\documentclass[preprint,showpacs,preprintnumbers,amsmath,amssymb]{revtex4}

\usepackage{graphicx}
\usepackage{dcolumn}
\usepackage{bm}
\usepackage{amsmath} 
\usepackage{color} 
\usepackage{CJK}
\def\v#1{\mbox{\boldmath $#1$}}

\begin{document}

\begin{CJK*} {GB} { } 
\title{Isothermal transport  of a near-critical binary  fluid mixture 
\\ through a capillary tube with the
preferential adsorption}

\author{Shunsuke Yabunaka}%
\email{yabunaka123@gmail.com}
\affiliation{Advanced Science Research Center, Japan Atomic Energy Agency, Tokai, 319-1195, Japan}

\author{Youhei Fujitani}
 \email{youhei@appi.keio.ac.jp}
\affiliation{School of Fundamental Science and Technology,
Keio University, 
Yokohama 223-8522, Japan}

\date{\today}

\begin{abstract}
We study isothermal transport of {a} binary fluid mixture, which lies in the homogeneous phase
near the demixing critical point, through a capillary tub{e}.  
{A short-range interaction is assumed between each mixture component and tube's wall surface, which
usually attracts one component more than the other. 
The resulting preferential} adsorption becomes significant owing to large osmotic susceptibility.
The mixture flowing out of the tube is rich in the preferred component when flow is driven by the pressure difference between the 
reservoirs. 
When flow is driven by the mass-fraction difference,
the total mass flow occurs in the presence of
 the preferential adsorption.
{T}hese phenomena
can be regarded as cross effects linked by the reciprocal relation. 
{The latter implies  that diffusioosmosis  
arises from the free energy of the bulk of the mixture not involving the surface potential, unlike 
usual diffusioosmosis far from the critical point.} 
We also study these phenomena numerically by using the hydrodynamics based on the coarse-grained free-energy functional, 
which was previously obtained to {reveal near-critical static properties}, {and using
material constants, which were previously obtained in some experimental studies.}
{Influence of the critical enhancement of the transport coefficients is found to be negligible
because of off-critical composition in the tube.} 
 It is {also} shown that
the conductance, or the total mass flow rate under a given
mass-fraction difference, 
can change non-monotonically with the temperature.
{The change is well expected to be large enough to be detected experimentally.} 
\end{abstract}

\maketitle
\end{CJK*}
\section{\label{sec:intro}Introduction}
{In this article, a binary fluid mixture containing no ions and lying}
in the homogeneous phase close to the demixing critical point is considered, {and is simply referred to as a mixture.}    
A short-range interaction, such as a dipole-dipole interaction, {is assumed} between each mixture component and 
a solid surface {in contact with a mixture.
Usually,} one component is {more} attracted by the surface {owing to}
difference in the interactions.  
The adsorption layer, where the preferred component is
more concentrated, can be of significant thicknes{s}
because o{f} large osmotic susceptibility {of a mixture} \cite{beysens1982, beysens1985, binder, cahn, bonn}.   
{Some static properties due to the 
preferential adsorption (PA) have been} studied
by using {t}he
renormalized local functional theory \cite{fisher-auyang,OkamotoOnuki}.
{At equilibrium, the composition profile significantly fluctuates on length scales smaller than the correlation length,
which can reach $100\ $nm experimentally.} 
After coarse-grained, many profiles that differ only on these scales are unified into much fewer profiles.
{In the theory,}
the {f}ree-energy functional, {coarse-grained up to the local correlation length,
is minimized by} the equilibrium profile, which 
{is made inhomogeneous by the PA. }\\

\noindent
We can apply hydrodynamics to study flow {of a mixture if the flow has} 
a typical length {l}arge enough in comparison to the correlation lengt{h} \cite{okafujiko, furu}.  
The hydrodynamics can be formulated {from} a coarse-grained free-energy functional \cite{onukibook}.
The inhomogeneity of the composition {profile}, linked with
that of the correlation length, \textcolor{black}{yields} 
 additional hydrodynamic stress.  The transport coefficients appearing in the hydrodynamics,
enhanced by \textcolor{black}{critical fluctuations} \cite{kawasaki,ohta}, are dependent on the local correlation length.  
Using the hydrodynamics based on the renormalized local functional theory \cite{yabuokaon, undul},
the present authors calculated
the drag coefficient of a colloidal particle in {a} mixture, which has 
the critical composition far from the particle \cite{yabufuji}.  
 The particle motion deforms the surrounding adsorption layer, which influences
the drag force.   
The inhomogeneity of the correlation length was found to be crucial for its dependence on the temperature.  \\

\begin{figure}
\includegraphics[width=6cm]{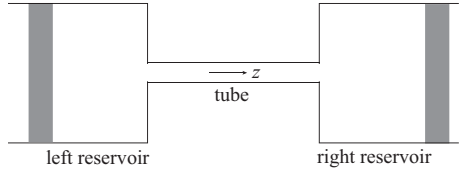}
\caption{{Schematic of a situation supposed in our formulation.
A} mixture is filled in the container composed of two reservoirs
and a tube connecting them. {The tube extends along the $z$ axis, with
the right reservoir lying on the positive side. Thick walls represent pistons.} }
\label{fig:tube}
\end{figure}

\noindent
We conside{r} isothermal transport of {a} mixture
 through {a} tube connecting two sufficiently large reservoir{s}.  
The transport is assumed to be caused by imposed differences in pressure and mass fraction 
 between the reservoirs, with the mixture remaining in the homogeneous phase. { 
One mixture component can be preferentially adsorbed by the tube wall.
A supposed situation is schematically drawn
in Fig.~\ref{fig:tube}.  A pressure difference is imposed by manipulating the pistons, whereas
{a} mass-fraction difference is imposed by setting mixtures with different composition{s} in the reservoirs.}
Applying the hydrodynamics for
a weak stationary laminar flow in the tube \cite{gromaz}, we study how the PA onto the tube wall
affects {the transport properties.}
The tube  is assumed to be so thin that the adsorption layer is not negligible
but is so thick that  the no-slip boundary condition can be imposed. 
The tube is assumed to be so long that 
effects of tube edges on the laminar flow and on the mixture in each reservoir are negligible.  In particular,
inhomogeneity of the correlation length is taken into account in our calculation,
unlike in the previous studies on similar isothermal transports \cite{wolynes, roij, xu}.  
\\

\noindent
Much attention has been paid to osmotic
 flow of a solution through a channel {\cite{osmflow, uematsu, keh, veleg, shin, marb, werk, sun, haq1, haq2, siva, frenkel}.}
A solution in the interfacial layer formed immediately near a channel wall can have distinct properties
due to some interaction potential between a solute and a {channel's} wall surface.
In a nonelectrolyte solution, the layer can be generated by the van der Waals interaction or some dipole--dipole interaction.
When a gradient of solute concentration is imposed in the channel direction, 
some force exerted on only the solution in the interfacial layer
can yield an apparent slip velocity
between the wall and the bulk of the solution, resulting in convection of the latter. 
This is the conventional explanation for diffusioosmosis {\cite{anders, derja}}.
In contrast, as shown later, when the mass-fraction difference is imposed in our problem,
diffusioosmosis {arises via} the anisotropic stress, {which does not} involve
any potential due to the tube's wall surface {but}
originates from the free energy in the bulk of a mixture.
{T}he length range over which each component interacts with the surface
is invisible in our coarse-grained description. The anisotropic stress 
varies inside the adsorption layer
distinctly thick in comparison with the molecular size.  
This mechanism is not considered in the conventional theory on diffusioosmosis,
{and was recently discussed as a possible mechanism for} diffusiophoresis {by one of the present authors} \cite{diffphore}.
{T}he {resulting} mass flow rates
should be sensitive to the temperature {owing to the critical singularity of the susceptibility}. 
 \\

\noindent
{In Section \ref{sec:form}},
we review the Onsager coefficients, formulate the
hydrodynamics, and describe the calculation procedure.  The pressure felt by the piston 
is equal to the negative of the grand-potential density in the central region of the reservoir;
this density depends not only on {the total mass density, denoted by $\rho$,} 
but also on {the composition}.
{We need not assume the mixture to be incompressible in the dynamics}.  
In Section \ref{sec:sol}, we derive some result{s}, neither using the renormalized local 
functional theory nor assuming that {the total mass density}
{i}s homogeneous at equilibriu{m.} 
{In Sections \ref{sec:pert} and \ref{sec:sol}, 
the cross-section perpendicular to a tube axis is assumed to be the same throughout the tube,
which need not be cylindrical.} We numerically calculate the Onsager coefficients and the mass flow rates
by using a specified model based on the renormalized local functional theory.
As mentioned in Section \ref{sec:ele},
it is assumed in this model that the tube is cylindrical, that $\rho$
is homogeneous at equilibrium, and that the mixture at equilibrium has the critical composition 
in the central region of the reservoir. 
In the numerical results {of} Section \ref{sec:num}, we
use material constants of a mixture of 2,6-lutidine and wate{r} and ones of a mixture of
nitroethane and 3-methylpentan{e}.  
The last section is devoted to a summary and further discussion. 
\section{Formulation\label{sec:form}}
{Mixture components are referred to as a and b;
$\rho_{n}$ denotes the mass density of the component ${n}(={\rm a\ or\ b})$.  
The total mass density $\rho$ equals $\rho_{\rm a}+\rho_{\rm b}$, 
whereas the difference $\rho_{\rm a}-\rho_{\rm b}$ is denoted by $\varphi$.}
We write $c_n$ for $\rho_n/\rho$ and
$\mu_{n}$ for the chemical potential conjugate to $\rho_{n}$.
 In the bulk of an equilibrium mixture, $\mu_{n}$ is a function of the
temperature (denoted by $T$), pressure ($P$), and the mass fraction of the component a ($c_{\rm a}$).
We write $\mu_\pm$ for  $\left(\mu_{\rm a}\pm\mu_{\rm b}\right)/2$,
and \textcolor{black}{define $\mu$} as the deviation of $\mu_-$ from
its critical value, $(\mu_{-})_{\rm c}$.  
The subscript $_{\rm c}$ generally indicates the value at the critical point;
$\rho$ and $\varphi$ are conjugate to $\mu_+$ and $\mu_-$, respectively.  
A quantity in each reservoir is indicated with the subscript $_{R}$ or $_{L}$;
${\cal M}_{{n} {\rm R}}$ denotes the total mass of the component ${n}$ in the right reservoir of Fig.~\ref{fig:tube}.
\subsection{Thermodynamics\label{sec:onsager}}
We first consider \textcolor{black}{entropy fluctuations of an equilibrium mixture}
in the isolated container with the pistons fixed (Fig.~\ref{fig:tube}).
Neglecting the contribution from the mixture in the tube,
we can regard the total entropy of the mixture in the container, denoted by $S$, as a function of
${\cal M}_{{\rm aR}}$, ${\cal M}_{{\rm bR}}$, 
and the internal energy of the mixture in the right reservoir, 
{because the total mass of each component and the total internal energy 
in the isolated container are constant.}
Writing $t$ for the time, we have \begin{equation}
 \frac{dS}{dt}={\sum_{{n}={\rm a, b}}\frac{\partial S}{\partial {\cal M}_{{n}{\rm R}}}\frac{d{\cal M}_{{n}{\rm R}}}{dt}}
=-\sum_{{n}={\rm a, b}} \frac{\mu_{n{\rm R}}-\mu_{n{\rm L}} }{{T}}\frac{d{\cal M}_{{n}{\rm R}}}{dt}
\label{eqn:DeltaS}\end{equation}
up to the second order of the magnitudes of the deviations.
Here, we drop the term {involving the internal energy, {which vanishes on the isothermal condition.}}
Equation (\ref{eqn:DeltaS}) {is included in} Eq.~(XV-55) of Ref.~\cite{gromaz},
{and tells that
the thermodynamic force{s}   are given by}
$-\left(\mu_{n{\rm R}}-\mu_{n{\rm L}}\right)/{T}$  and the conjugate flux{es} by
$d{\cal M}_{n{\rm R}}/(dt)$, {with ${n}$ being 
a and b,}  in the isothermal transport.
\\

\noindent
Imposing weak thermodynamic forces on an equilibrium state \textcolor{black}{throughout the mixture}, we 
consider a weak \textcolor{black}{flow} in terms of the {linear} nonequilibrium thermodynamics.
The mass densities are homogeneous in the central region of each reservoir;
a superscript $^{({\rm ref})}$ is added to a quantity in this region in the equilibrium state.
For example, the value of $c_{n}$ shared by the central regions in the reservoirs in the equilibrium state
is referred to as $c_n^{({\rm ref})}$, which is \textcolor{black}{here} assumed 
to be \textcolor{black}{either equal or} close to \textcolor{black}{the value at the critical composition}, $c_{n{\rm c}}$.
A flow is driven by a nonzero value of $\delta c_{\rm a}$, 
{which is defined as 
the difference of the value of $c_{\rm a}$
in the central region of the left reservoir subtracted from that of the right reservoir. 
In this manner,} $\delta$ is used to indicate the difference \textcolor{black}{in a quantity} {between these regions.}
An equilibrium mixture has homogeneous chemical potentials.
Even in the presence of flow in the tube,
the mixture \textcolor{black}{is} considered to be at  equilibrium in the large reservoir {s}.
The difference in the numerator in Eq.~(\ref{eqn:DeltaS}) can be written as
$\delta \mu_n$.
In each reservoir, because of the mechanical balance,  the pressure felt by the \textcolor{black}{right (left)} piston in Fig.~\ref{fig:tube}
is always the same as the pressure in the central region of the \textcolor{black}{right (left)} reservoir.   
As a result, $\delta P$ equals the difference {between}
{the} pressures felt by the pistons
{on both sides} and {can} be controlled externally.
\\

\noindent
{W}e consider another set of  forces,
$-\delta P/T$ and $-\delta\mu/T$.  Neglecting  possible 
\textcolor{black}{effects of the tube edges}, {we assume
that $P$ and $\mu$  are constant on the cross section at 
each tube edge to regard
$\delta P$ and $\delta \mu$
as the respective} differences between the tube edges.
With the aid of the Gibbs-Duhem relation for $\delta T=0$,
we have
\begin{equation}
\left(\begin{array}{cc}{\delta P}\\ \delta \mu\end{array}\right)=\left(\begin{array}{cc}
\rho_{\rm a}^{({\rm ref})} &\rho_{\rm b}^{({\rm ref})} 
\\1/2&-1/2\end{array} \right)\left(\begin{array}{cc}{\delta \mu}_{\rm a}\\ \delta \mu_{\rm b}\end{array}\right)
\ .\end{equation}
{Using the coefficient matrix above, denoted by $\Theta$,
we convert the fluxes in Eq.~(\ref{eqn:DeltaS})
to the fluxes conjugate to $(-\delta P/T, -\delta\mu/T)$.  
The fluxes, $\left({\cal I}, {\cal J}\right)$, are defined}  
by the first equality of
\begin{equation}
\left[\Theta^{-1}\right]^{\rm T}
\left(\begin{array}{c} d{{\cal M}_{\rm aR}}/(dt)  \\ d{{\cal M}_{\rm bR}}/(dt) 
\end{array}\right)= \left(\begin{array}{c} {\cal I} \\ {\cal J} \end{array}\right)=
{\cal L}
\left(\begin{array}{c}  -\delta P/{T} \\ -\delta\mu/{T} \end{array}\right)
\ ,\label{eqn:newlinph}\end{equation}
where the superscript $^{\rm T}$ indicates the transposition.  The second equality above 
represents the linear phenomenological equations (LPEs) \cite{ons1,ons2}.
{The components of the coefficient matrix ${\cal L}$ are Onsager coefficients, 
satisfying  the reciprocal relation,
${\cal L}_{12}={\cal L}_{21}$, with the matrix being positive semi-definite.} 
Using the first equality of Eq.~(\ref{eqn:newlinph}), we have
\begin{equation}
\frac{d}{dt} {\cal M}_{n{\rm R}}=\rho_n^{({\rm ref})}{\cal I}\pm \frac{1}{2}{\cal J}
\ ,\label{eqn:marmbr}\end{equation}
where the upper (lower) sign is taken for $n=$a (b) in the double sign.
{The total mass flow rate is denoted by $d{\cal M}_{{\rm R}}/(dt)$ and equals
the sum of Eq.~(\ref{eqn:marmbr}) over $n=$ a and b, {\it i.e.}\ , $\rho^{({\rm ref})} {\cal I}$. }
\\

\noindent
We later discuss flow driven by the pressure \textcolor{black}{difference}  
$(\delta P\ne 0\ ,\delta c_{\rm a}=0)$ and flow driven by the mass-fraction difference $(\delta P=0\ ,\delta c_{\rm a}\ne 0)$;
$\delta c_{\rm a}$ would be easier to handle experimentally than $\delta \mu$.
The partial volume per unit mass of the component $n$, denoted by ${\bar v}_n$, is defined as
the partial derivative of the mixture volume with respect to the mass of the component $n$
with $T$, $P$, and the mass of the other component held constant.
We write ${\bar v}_\pm$ for $({\bar v}_{\rm a}\pm{\bar v}_{\rm b})/2$.
The volume and masses are extensive variables, which leads to
\begin{equation}
1 = {\bar v}_{\rm a} \rho_{\rm a}+ {\bar v}_{\rm b} \rho_{\rm b}
={\bar v}_+\rho  + {\bar v}_-\varphi
\label{eqn:rhovv}\end{equation}
in the bulk of an equilibrium mixture. 
The composition can be represented by either $c_{\rm a}$ or ${\varphi}{\equiv \rho(2c_{\rm a}-1)}$.
{F}or the isothermal transport, {w}e have
\begin{equation}
\delta \mu_n=
\left.\frac{\partial \mu_n }{\partial P}\right)_{T, c_{\rm a}}\ \delta P
+\left.\frac{\partial \mu_n }{\partial c_{\rm a}}\right)_{T, P}\ \delta c_{\rm a}
\ ,\label{eqn:delmua}\end{equation}
{where}  
the fixed variables in the partial differentiation are indicated in the subscript of a right parenthesis and
the partial derivatives {are} evaluated at the central region of the reservoir 
in the equilibrium state.
{The first derivative on the right-hand side (RHS) above
equals ${\bar v}_n^{({\rm ref})}$ owing to} a Maxwell relation {derived from the Gibbs free energy.}
Thus, we have
\begin{equation}
\delta \mu= {\bar v}_-^{({\rm ref})}\delta P +\left.\frac{\partial \mu }{\partial c_{\rm a}}\right)_{T, P}\ \delta c_{\rm a}\label{eqn:kakkoP2}
\ .\end{equation}
The second term on {the RHS} above represents the part of $\delta\mu$
caused by $\delta c_{\rm a}$ and is below denoted by $\left(\delta \mu\right)_P$.
{The partial derivative in the second term 
is} positive owing to the thermodynamic stability, is proportional to the inverse of the osmotic susceptibility, 
as shown in Eq.~(\ref{eqn:fromdiffo}) below, and
becomes closer to zero as the critical point is approached.
Using Eq.~(\ref{eqn:kakkoP2}), we can convert the thermodynamic forces
{to}  $\left(-\delta P/T,
-\left(\delta\mu\right)_P/T\right)$.   
\subsection{Hydrodynamics\label{sec:hyd}}  
{W}e assume the free energy of the bulk of a mixture to be given by the volume integral of a function
of $\rho_{\rm a}$, $\rho_{\rm b}$ and the quadratic form of their gradients
over the mixture region, $V_{\rm tot}$.
To describe the PA, we add
the {area} integral of a function of $(\rho_{\rm a}, \rho_{\rm b})$, or $(\rho, \varphi)$, over the interface between
the mixture and the container, $\partial V_{\rm tot}$.
With $f_{\rm bulk}$ and $f_{\rm surf}$ denoting these functions, respectively, the free-energy functional
is given by
\begin{equation}
{F}[\rho_{\rm a}, \rho_{\rm b}]=\int_{V_{\rm tot}}d{\v r}\ 
f_{\rm bulk} \left(\rho, \varphi, \nabla\rho, \nabla\varphi \right)
+ \int_{\partial V_{\rm tot}}dA\ f_{\rm surf}\left(\rho, \varphi\right)
\ ,\label{eqn:general}\end{equation}
\textcolor{black}{where} the mass densities depend on the position ${\v r}$. 
The dependence of $f_{\rm bulk}$ on the gradients is assumed to be only via
$\left|\nabla\rho\right|^2, \left|\nabla\varphi\right|^2$, and $(\nabla\rho)\cdot (\nabla\varphi)$.
Because Eq.~(\ref{eqn:general}) is the Helmholtz free energy of a mixture with
$\rho_{\rm a}({\v r})$ and $\rho_{\rm b}({\v r})$ being given,  
$\mu_n({\v r})$ is given by  the functional derivative of  ${F}[\rho_{\rm a},\rho_{\rm b}]$ 
with respect to $\rho_n({\v r})$
in the bulk of a mixture, which leads to
\begin{equation}
\mu_+= \frac{\partial f_{\rm bulk}}{\partial \rho}-\nabla\cdot\left( \frac{\partial f_{\rm bulk}}{\partial \nabla\rho}\right)
\quad{\rm and}\quad
 \mu_-= \frac{\partial f_{\rm bulk}}{\partial \varphi}-\nabla\cdot\left( \frac{\partial f_{\rm bulk}}{\partial \nabla\varphi}\right)    
\ .\label{eqn:murhophi}\end{equation}
{W}e can obtain the reversible part of the pressure tensor, $\Pi_{\rm rev}$, by considering
\textcolor{black}{how} the free energy is changed by a quasistatic deformation of a mixture.
Its scalar part, 
$P$, equals the negative of the grand-potential density, {\it i.e.\/},
\begin{equation}
P=-f_{\rm bulk}+\rho_{{\rm a}}\mu_{\rm a}+\rho_{{\rm b}}\mu_{\rm b}=
-f_{\rm bulk}+\rho\mu_++\varphi\mu_-
\label{eqn:Pequal}
\ .\end{equation}
As shown in Appendix \ref{sec:stress},  with ${\v 1}$ denoting {the identity tensor of order two},
we have 
\begin{equation}
\Pi_{\rm rev}
=P{\v 1}+\frac{\partial f_{\rm bulk}}{\partial \left(\nabla\rho\right)} \left(\nabla\rho\right)
+\frac{\partial f_{\rm bulk}}{\partial \left(\nabla\varphi\right)} \left(\nabla\varphi\right)
\ ,\label{eqn:genPi}\end{equation}
which is a symmetric tensor. 
After some algebra, we obtain
\begin{equation}
\nabla \cdot \Pi_{\rm rev}
=\rho\nabla \mu_++\varphi\nabla\mu
\ .\label{eqn:Piosmgen}\end{equation}
{For quantities in the central regions of the reservoirs, w}e use Eqs.~(\ref{eqn:murhophi}) and (\ref{eqn:Pequal}) to find 
\begin{equation}
\delta P=\rho^{({\rm ref})}\delta \mu_+ +\varphi^{({\rm ref})} \delta \mu
\ .\label{eqn:delPgen}\end{equation}
\textcolor{black}{Thus,} considering Eq.~(\ref{eqn:kakkoP2}), $\delta \mu_+$ is determined by $\delta P$ and $\delta c_{\rm a}$.
\\

\noindent
The diffusion {flux} of the component {${n}(= {\rm a\ or\ b\ })$}, denoted by {${\v j}_n$, 
is defined so that the sum ${\v j}_{\rm a}+{\v j}_{\rm b}$} vanishes. We have
\begin{equation}
\frac{\partial \rho_{n}}{\partial t}=-\nabla\cdot \left( \rho_{n} {\v v}\right) -\nabla\cdot {\v j}_{n} 
\label{eqn:alpdiffusion}\end{equation} 
{from} the mass conservation. The sum {of the above} over $n=$ a and b is
\begin{equation}
\frac{\partial \rho}{\partial t}= -\nabla\cdot \left(\rho{\v v}\right)
\ ,\label{eqn:renzoku}\end{equation}
whereas the difference gives
\begin{equation}
\frac{\partial \varphi}{\partial t}=-\nabla\cdot \left( \varphi {\v v}\right) -\nabla\cdot \textcolor{black}{{\v j} 
\ ,}\label{eqn:diffusion}\end{equation}
{where ${\v j}$ is defined as}  
${\v j}_{\rm a}-{\v j}_{\rm b}$.  
The left-hand sides (LHSs) of Eqs.~(\ref{eqn:alpdiffusion}) {--} (\ref{eqn:diffusion})
vanish in the stationary state.   
We can assume ${\v j}$ to be equal to $-\Lambda \nabla\mu$, where
$\Lambda$ denotes the transport coefficient
\textcolor{black}{of} the interdiffusion.
The container's wall surface is assumed to be impermeable to the mixture, which means that
the normal component of $\nabla\mu$ vanishes at $\partial V_{\rm tot}$.  
As mentioned later, 
in a weak stationary laminar flow in the tube, 
$\nabla\cdot {\v v}$ vanishes
although $\rho$ is not assumed to be constant.  \textcolor{black}{There,} the momentum conservation gives 
\begin{equation}
0 =-\rho\nabla \mu_+-\varphi\nabla\mu+2\nabla\cdot\left(\eta_{\rm s} E\right)
\ ,\label{eqn:stok}
\end{equation}
where $\eta_{\rm s}$ is the shear viscosity
and $E$ is the rate-of-strain tensor.  
The no-slip boundary condition is imposed at $\partial V_{\rm tot}$.
{T}he transport coefficients, $\eta_{\rm s}$ and $\Lambda$, can be inhomogeneous.
\\

\noindent
Because the fields are coarse-grained up to the local correlation length,
the mass densities at equilibrium minimize the grand potential of a mixture,
{\it i.e.\/}, 
\begin{equation} {F}[\rho_{\rm a}, \rho_{\rm b}]
- \int_{V_{\rm tot}}d{\v r} \ \left[\mu_+^{({\rm ref})} \rho({\v r})+\mu_-^{({\rm ref})} \varphi({\v r})\right]
\ .\label{eqn:relation}\end{equation}  
 Thus, the equilibrium mass densities satisfy the two equations in
Eq.~(\ref{eqn:murhophi}) with the superscript $^{({\rm ref})}$ added to the chemical potentials, together with the boundary conditions
\begin{equation}
{\v n}_{\partial V_{\rm tot}}\cdot \frac{\partial f_{\rm bulk}}{\partial \left(\nabla \rho\right)}+
\frac{\partial f_{\rm surf}}{\partial \rho}={\v n}_{\partial V_{\rm tot}}\cdot \frac{\partial f_{\rm bulk}}{\partial \left(\nabla \varphi\right)}+
\frac{\partial f_{\rm surf}}{\partial \varphi}=0
\quad{\rm at}\ \partial V_{\rm tot}
\ ,\label{eqn:surfbound}\end{equation}
where ${\v n}_{\partial V_{\rm tot}}$ denotes the outward facing unit vector normal to $\partial V_{\rm tot}$.  
Homogeneous chemical potentials imply the mechanical balance \textcolor{black}{at} equilibrium,  $\nabla\cdot\Pi_{\rm rev}=0$,
because of Eq.~(\ref{eqn:Piosmgen}).  Equation (\ref{eqn:rhovv}) need not hold  
in a region with inhomogeneous mass densities.
\subsection{Perturbation scheme\label{sec:pert}}
{A}ssuming that $\delta P$ and $\delta c_{\rm a}$
are proportional to a dimensionless smallness parameter, $\varepsilon$,
we calculate the fields in the tube up to the order of $\varepsilon$ to obtain the Onsager coefficients.
The superscripts $^{(0)}$ and $^{(1)}$ are used to indicate the order of $\varepsilon$.
For example, we have
$\mu_\pm=\mu_\pm^{(0)}+\varepsilon \mu_\pm^{(1)}$ up to the order of $\varepsilon$,
 where $\mu_\pm^{(0)}$ respectively equal $\mu_\pm^{({\rm ref})}$.
The tube is assumed to have the same cross section along the $z$ axis with
the right reservoir lying on the positive side {(Fig.~\ref{fig:tube})}. 
Taking the Cartesian coordinates $(x,y,z)$, 
we can assume 
$\rho^{(0)}$ and $\varphi^{(0)}$ to be functions of $x$ and $y$, which are coordinates on a cross section.
{Using} the area integral over a cross section of the tube, $S_{\rm tube}$, we have  
\begin{equation}
\frac{d}{dt}{\cal M}_{{n} {\rm R}}=\varepsilon \int_{S_{\rm tube}} dA\  \left(\rho_{n}^{(0)} v_z^{(1)} +j_{{n}z}^{(1)}\right)
\label{eqn:dMdt}\end{equation}
{u}p to the order of $\varepsilon$.
With the aid of the first equality of Eq.~(\ref{eqn:newlinph}), 
Eq.~(\ref{eqn:dMdt}) gives
\begin{eqnarray}
&&{\cal I}=\frac{\varepsilon}{\rho^{({\rm ref})}}  \int_{S_{\rm tube}}dA\ 
\rho^{(0)}v_z^{(1)}\label{eqn:IJexpress2a}\\
&& {\rm and}\quad {\cal J}=\varepsilon  \int_{S_{\rm tube}}dA\ \left[\left(\varphi^{(0)}-\frac{\rho^{(0)}\varphi^{({\rm ref})}}{\rho^{({\rm ref})}}\right)
v_z^{(1)}+j_z^{(1)}\right]\ .
\label{eqn:IJexpress2b}\end{eqnarray}
The transport coefficients {$\eta_{\rm s}$ and $\Lambda$} can be {regarded as} 
dependent on $\rho^{(0)}$ and $\varphi^{(0)}$ in the {e}quations up to the order of $\varepsilon$.  
The resulting $(x, y)$-dependent {coefficients}
are denoted by $\eta_0(x, y)$ and $\Lambda_0(x, y)$, respectively.  
In the absence of PA, 
{the mass densities become homogeneous and
Eq.~(\ref{eqn:stok}) becomes the usual Stokes equation, $0=-\nabla P+\eta_{\rm s}\Delta{\v v}$.}
In Appendix  \ref{sec:insep}, 
our formulation up to here is shown to be consistent with
the reciprocal relatio{n}.
\section{Immediate results from the formulation\label{sec:sol}}  
Because \textcolor{black}{${\v v}$ vanishes for $\varepsilon=0$,}   
Eq.~(\ref{eqn:renzoku}) in a stationary laminar flow in the tube gives 
\begin{equation}
0=\nabla\cdot {\v v}^{(1)}=\partial_z v_z^{(1)}
\ ,\label{eqn:nablavtube}\end{equation}
where $\partial_z$ denotes the partial derivative with respect to $z$.  {We obtain}
\begin{equation}
0=\rho^{(0)}{\bar\nabla} \mu_+^{(1)}+\varphi^{(0)} {\bar\nabla} \mu^{(1)}
\label{eqn:ppvpmxy}\end{equation} 
{from} the $x$ and $y$ components of Eq.~(\ref{eqn:stok}). Here,
${\bar \nabla}$ represents the two-dimensional nabla defined on the $(x, y)$ plane.
{The} $z$ component gives
 \begin{equation}
{\bar \nabla}\cdot \left({\eta}_0 {\bar \nabla} v_z^{(1)}\right)
=\rho^{(0)}\partial_z\mu_+^{(1)}+\varphi^{(0)}\partial_z \mu^{(1)}
\ ,\label{eqn:st3xy}\end{equation}
where $v_z^{(1)}$ is regarded as a scalar on the plane.   The LHS above
is independent of $z$ {because of Eq. (\ref{eqn:nablavtube})}.
In a stationary laminar flow, Eq.~(\ref{eqn:diffusion}) yields $0=\nabla\cdot\left({\Lambda_0} \nabla \mu^{(1)}\right)$, {\it i.e.\/},
\begin{equation}
0={\bar \nabla}\cdot \left( \Lambda_0 {\bar \nabla} \mu^{(1)}\right)
+\Lambda_0\partial_z^2 \mu^{(1)} \ .\label{eqn:diffusion2}
\end{equation}
The boundary condition of ${\bar \nabla}\mu^{(1)}$ on a cross section is given below
Eq.~(\ref{eqn:diffusion}).  As assumed in the third paragraph of \textcolor{black}{Section} \ref{sec:onsager},
$\mu$ is constant over the cross section at each of the tube edges.  
These conditions and Eq.~(\ref{eqn:diffusion2}) are satisfied when
$\mu^{(1)}$ is a linear function of $z$ \textcolor{black}{and} is independent of $x$ and $y$.
\textcolor{black}{Then,} because of Eqs.~(\ref{eqn:ppvpmxy}) and (\ref{eqn:st3xy}), 
$\mu_+^{(1)}$ is also a linear function of $z$ and is independent of $x$ and $y$.
Hence, with the aid of Eq.~(\ref{eqn:delPgen}), we obtain
\begin{equation}\varepsilon
\partial_z \mu_+^{(1)}= \frac{\delta P-\varphi^{({\rm ref})}\delta \mu}{\rho^{({\rm ref})}L_{\rm tube}} \quad {\rm and}\quad
\varepsilon\partial_z \mu^{(1)}= \frac{\delta\mu}{L_{\rm tube}} 
\ ,\label{eqn:nablamuP}\end{equation}
where $L_{\rm tube}$ denotes the length of the tube.  Thus, Eq.~(\ref{eqn:st3xy}) becomes
\begin{equation}
\varepsilon {\bar \nabla}\cdot \left({\eta}_0 {\bar \nabla} v_z^{(1)}\right)
=\frac{\rho^{(0)}\delta P}{\rho^{({\rm ref})}L_{\rm tube}}+ \left( \varphi^{(0)}-\frac{\rho^{(0)}\varphi^{({\rm ref})}}{\rho^{({\rm ref})}} 
\right)\frac{\delta\mu}{L_{\rm tube}}
\ .\label{eqn:st3xy+}\end{equation}
Owing to ${\v j}^{(1)}=-\Lambda_0\nabla\mu^{(1)}$, ${\v j}^{(1)}$, {and hence} 
${\v j}_{\rm a}^{(1)}=-{\v j}_{\rm b}^{(1)}$, {are} along the $z$ axis. 
{This yields}
\begin{equation}
\varepsilon j_z^{(1)}\textcolor{black}{(x,y)}= -\Lambda_0\textcolor{black}{(x,y)}\frac{\delta \mu }{L_{\rm tube}}\ .
\label{eqn:jzprofile}\end{equation}
In the absence of PA, 
the difference in the parentheses on the RHS of Eq.~(\ref{eqn:st3xy+}) identically vanishes,  and 
$v_z^{(1)}$ is independent of $\delta \mu$.
That the difference identically vanishes is equivalent with that $c_{\rm a}^{(0)}(x, y)$ equals $c_{\rm a}^{({\rm ref})}$ for any $(x, y)$.
This equality should mean no PA, and {hence} 
we can assume that the difference identically vanishes only in the absence of PA.
\\

\noindent
In the presence of PA, because $v_z^{(1)}$ satisfies Eq.~(\ref{eqn:st3xy+}) and the no-slip boundary condition at
the tube wall, the solution for $v_z^{(1)}$ is the sum of a term proportional to $\delta P$ and a term proportional 
to $\delta\mu$.
Substituting this solution and Eq.~(\ref{eqn:jzprofile}) into Eqs.~(\ref{eqn:IJexpress2a}) and (\ref{eqn:IJexpress2b}),
we should be able to express the Onsager coefficients ${\cal L}_{ij}$, appearing in the LPEs of Eq.~(\ref{eqn:newlinph}),
in terms of the quantities at the order of $\varepsilon^0$. 
Although the solution is not obtained at this stage, 
we find that ${\cal L}_{11}$ and ${\cal L}_{12}={\cal L}_{21}$ involve $v_z^{(1)}${,} {\it i.e.\/}, convection.
Part of the first term in the brackets 
of  Eq.~(\ref{eqn:IJexpress2b}) contributes to ${\cal L}_{22}$ and
the second term also contributes to ${\cal L}_{22}$.
We refer to the parts of ${\cal L}_{22}$ produced by these contributions as ${\cal L}_{22{\rm v}}$ and ${\cal L}_{22{\rm d}}$, respectively.
The former involves convection, whereas the latter interdiffusion.  {Neither of them contributes to the total mass flow.}
Notably, $\rho^{(1)}$ and $\varphi^{(1)}$
can be determined by Eq.~(\ref{eqn:murhophi}) 
once $\mu_\pm^{(1)}$ are given, but we need not know $\rho^{(1)}$ and $\varphi^{(1)}$
in calculating ${\cal L}_{ij}$.
\\

\noindent
In the absence of PA, because  
${\cal L}_{12}={\cal L}_{21}$ and ${\cal L}_{22{\rm v}}$ {are found to} vanish, 
the first term on the RHS of Eq.~(\ref{eqn:marmbr}) is \textcolor{black}{equal} to 
$-\rho_n^{({\rm ref})} {\cal L}_{11} \delta P/T$ involving the convection, \textcolor{black}{whereas} 
the second term to $\mp {\cal L}_{22{\rm d}}\delta\mu/(2T)$
\textcolor{black}{involving} the interdiffusion.   {This means that,} 
when \textcolor{black}{f}low is driven by the pressure difference, {\it i.e.\/}, by $\delta P \ne 0$ and $\delta c_{\rm a}=0$,
the convected part of a mixture has the mass fraction of $c_n^{({\rm ref})}$.
Then, unless ${\bar v}_-^{({\rm ref})}$ vanishes in Eq.~(\ref{eqn:kakkoP2}), the \textcolor{black}{interdiffusion occurs in addition.
In} the absence, only the interdiffusion occurs and the total mass flow rate disappears when 
\textcolor{black}{f}low is driven by the mass-fraction difference, 
{\it i.e.\/}, by $\delta c_{\rm a}\ne 0$ and $\delta P=0$.  \\

\noindent
In the presence of PA, 
$-\rho_n^{({\rm ref})} {\cal L}_{12}\delta\mu/T$ \textcolor{black}{emerges} in the first term on the RHS of Eq.~(\ref{eqn:marmbr}), 
whereas $\mp {\cal L}_{21}\delta P/(2T)$ and $\mp {\cal L}_{22{\rm v}}\delta\mu/(2T)$ \textcolor{black}{emerge} in the second term.
Because of the second of these three emergent terms, when \textcolor{black}{f}low is driven by the pressure difference,
even the convected part of a mixture does not have the mass fraction of 
$c_n^{({\rm ref})}$, which is realized considering that the preferred component is concentrated in the tube. 
Because  of the first and third of the emergent terms, not only the interdiffusion but also the convective transport occurs
when flow is driven by the mass-fraction difference.
In particular,  the first term, involving ${\cal L}_{12}$, generates the total mass flow, which represents
diffusioosmosis.
The coefficients ${\cal L}_{21}$ and ${\cal L}_{12}$ represent the cross effects in 
the LPEs of Eq.~{(\ref{eqn:newlinph})}.
\section{Elements for numerical calculations\label{sec:ele}}
{Below}, we assume that a mixture, at equilibrium throughout,  
has the critical composition in the central region 
of the reservoir.
In other words, we assume $\rho_n^{({\rm ref})}=\rho_{n{\rm c}}$, which leads to 
$c_n^{({\rm ref})}=c_{n{\rm c}}$, $\rho^{({\rm ref})}=\rho_{\rm c}$, and
$\varphi^{({\rm ref})}=\varphi_{\rm c}$.
The difference $\varphi({\v r})-\varphi_{\rm c}$, denoted by $\psi({\v r})$, 
 is the order parameter of the phase separation. {In the absence of PA ($h=0$), {$\psi({\v r})$} vanishes.}
\subsection{Free-energy functional\label{sec:free}}
We assume that
$f_{\rm bulk}$ is separated into the $\rho$-dependent part and $\psi$-dependent part.  
The former part is a function of $\rho$, denoted by $f_+(\rho)$, whereas  
the latter part consists of a function of $\psi$, denoted by $f_-(\psi)$, and
the square gradient term.
This term is written as
$M_-(\psi) \left| \nabla\psi \right|^2/2$ with $M_-(\psi)$ denoting a positive function of $\psi$.
The expressions of  $f_-$ and $M_-$ are given in terms of the renormalized local functional theory \cite{fisher-auyang,OkamotoOnuki}, 
as shown in the next subsection. 
In this theory, as mentioned in Appendix \ref{sec:rlft}, 
the functional is obtained by coarse-graining {a} bare model
{up to the local correlation length of composition fluctuations},
and assuming $\varphi^{({\rm ref})}=\varphi_{\rm c}$ amounts to assuming
$\mu_-^{(0)}=\left(\mu_-\right)_{\rm c}$, {\it i.e.\/}, $\mu^{(0)}=0$.
{The correlation length is required to be much larger than a molecular size
for the coarse-grained description to be valid.}
We also assume that $f_{\rm surf}$, being independent of $\rho$,
is proportional to $\psi$ apart from an irrelevant constant.
The negative of the constant of proportionality, denoted by $h$, is called
the surface field \cite{cahn, bray, diehl86, diehl97}.  
On these assumptions, Eq.~(\ref{eqn:general}) becomes
\begin{equation}
{F}[\rho_{\rm a}, \rho_{\rm b}]=\int_{V_{\rm tot}}d{\v r}\ 
\left[ f_+(\rho) +f_-(\psi) +\frac{1}{2} M_-(\psi) \left\vert \nabla\psi \right\vert^2\right]
-h \int_{\partial V_{\rm tot}}dA\ \psi \ .\label{eqn:special}\end{equation}
Using the procedure mentioned in the last paragraph of Section \ref{sec:hyd}, 
{we can calculate $\rho^{(0)}$ and $\psi^{(0)}$.  In particular,} we find
that $\rho^{(0)}$ is homogeneously equal to $\rho^{({\rm ref})}=\rho_{\rm c}$.
The equilibrium profile of the order parameter between the two parallel plates and the one around a sphere
are calculated {using the renormalized local functional theory}
\cite{fisher-auyang,OkamotoOnuki, okaonPRE, Yabu-On, yabufuji}.  \\

\noindent
Because $\rho^{(0)}$ is homogeneous,  
Eq.~(\ref{eqn:renzoku}) at the order of $\varepsilon$ gives
$\nabla\cdot {\v v}^{(1)}=0$ in a stationary flow, whether
it is laminar or not.  Equation (\ref{eqn:murhophi}) gives $\mu_+=f_+'(\rho)${, which yields}
$\mu_+^{(1)}=\rho^{(1)} f_+''(\rho^{(0)})$. 
We define $p$ as $\rho f_+'-f_+$  to obtain $p^{(1)}=\rho^{(0)}\mu_+^{(1)}$, 
and Eq.~(\ref{eqn:Piosmgen}) gives 
\begin{equation} 
\nabla\cdot\Pi^{(1)}_{\rm rev}=
\nabla p^{(1)}+\varphi^{(0)}\nabla \mu^{(1)}\ ,
\label{eqn:Piosmgen1}\end{equation}
which is used in Ref.~\cite{yabufuji}.  
\textcolor{black}{The first term on the RHS of Eq.~(\ref{eqn:ppvpmxy}) can be replaced by ${\bar\nabla} p^{(1)}$, whereas
that of Eq.~(\ref{eqn:st3xy}) by $\partial_z p^{(1)}$.
I}n Eq.~(\ref{eqn:IJexpress2a}), 
${\cal I}$ becomes equal to the area integral of $\varepsilon v_z^{(1)}$ over ${\rm S}_{\rm tube}$ {and
gives the flow rate.} 
The difference in the parentheses of Eq.~(\ref{eqn:IJexpress2b}) equals $\psi^{(0)}$. 
{E}ven if ${\cal I}$ vanishes, the diffusion fluxes can change the composition of 
each reservoir, {and hence} its volume. 
{Equations} (\ref{eqn:Pequal}) {and} (\ref{eqn:delPgen}) give
$P=p-f_-+\varphi\mu_-$ and  
\begin{equation}
\delta P=\delta p+\varphi_c\delta \mu\ ,
\label{eqn:dP}\end{equation}
{respectively. I}f the mass densities are homogeneous, we have
\begin{equation}
\left.\frac{\partial\mu}{\partial \varphi}\right)_{T\rho}= \left.\frac{\partial\mu}{\partial P}\right)_{Tc_{\rm a}}
 \left.\frac{\partial P}{\partial \varphi}\right)_{T\rho} 
+\left.\frac{\partial\mu}{\partial c_{\rm a}}\right)_{TP}
 \left.\frac{\partial c_{\rm a}}{\partial \varphi}\right)_{T\rho}
\ .\end{equation}
The LHS above equals \textcolor{black}{the inverse of the osmotic susceptibility,} 
$f_-''(\psi)$, whereas
the first and second derivatives of the first term on the RHS 
equal ${\bar v}_-$ and $f_-''(\psi)\varphi$, respectively. 
{We can apply
Eq. (\ref{eqn:rhovv}) and ${\varphi}{\equiv \rho(2c_{\rm a}-1)}$
for the second derivative of the second term.  Thus,  in}
the central region of the reservoir at equilibrium throughout the mixture, we obtain
 \begin{equation}
 \left.\frac{\partial\mu}{\partial c_{\rm a}}\right)_{TP} =2\rho_{\rm c}^2{\bar v}_{+}^{({\rm ref})} f_{-}''(0)\ ,\label{eqn:fromdiffo}
\end{equation}
which can be substituted into Eq.~(\ref{eqn:kakkoP2}). {A}s shown by Eq.~(\ref{eqn:fprpr}) below,
{we have} $f_{-}''(0){\propto} \tau^\gamma$. 
\subsection{Non-dimensionalization\label{sec:nond}}
Using the conventional notation, we write $\beta,\gamma,\nu,$ and $\eta$ for the critical exponents 
 {of} binary mixtures near the demixing critical point. 
 The exponent $\eta$ represents the deviation 
from the straightforward dimensional analysis of the equal-time 
correlation function of the order-parameter fluctuation\textcolor{black}{s} at the critical point. 
The critical temperature is denoted by $T_{\rm c}$,
and the reduced temperature $\tau$ is defined as $\left\vert T-T_{\rm c}\right\vert/T_{\rm c}$.
In an equilibrium mixture with the critical \textcolor{black}{composition} \textcolor{black}{($\psi=0$)}, 
the correlation length, denoted by $\xi$, is given by $\xi_0\tau^{-\nu}$
\textcolor{black}{in the critical regime}, where $\xi_0$ is a material constant.
In general, $\xi$ depends on \textcolor{black}{$\tau$ and $\psi$}.
The ``distance'' from the critical point is represented by $\omega$, which is defined so that we have
\begin{equation}
\xi=\xi_0 \omega^{-\nu}\ .\label{eqn:xiomega}\end{equation}
The values of the critical exponents are shown in Ref.~\cite{peli}; we adopt
$\nu= 0.630$ and $\eta= 0.0364$. The (hyper)scaling relations give $2\beta+\gamma=3\nu$ and
$\gamma=\nu(2-\eta)$.
Hereafter, the tube is assumed to be a cylinder with the radius $r_{\rm tube}$.  With $r$ denoting the radial distance
from the center axis, \textcolor{black}{we can write $\psi^{(0)}(r), \eta_0(r),$ and $\Lambda_0(r)$ because they are functions of $r$.}
\\

\noindent
{A} dimensionless radial distance ${\hat r}$ is defined a{s} $r/r_{\rm tube}$.
 A characteristic reduced temperature $\tau_*$ is defined so that $\xi$ becomes $r_{\rm tube}$ for $\psi=0$ at $\tau=\tau_*$.  
A characteristic order parameter $\psi_*$ is defined so that
$\xi$ becomes  $r_{\rm tube}$  for $\psi=\psi_*$ 
at $\tau=0$.  According to {the self-consistent condition mentioned in Appendix \ref{sec:rlft},}
these two definitions are equivalent to 
\begin{equation}
\tau_*=\left(\frac{\xi_{0}}{r_{\rm tube}}\right)^{1/\nu}\quad {\rm and}\quad 
\psi_*=\frac{\tau_*^\beta}{\sqrt{C_2}}
\ .\label{eqn:nodim}\end{equation}
{Here,} $C_2$ is a material constant, {whose estimates are obtained
in Appendix \ref{sec:rlft}.}
{W}e use ${\hat \tau}\equiv \tau/\tau_*$, ${\hat\omega}\equiv \omega/\tau_*$, and 
${\hat \psi}^{(0)}({\hat r})\equiv \psi^{(0)}(r_{\rm tube}{\hat r} )/\psi_*$, {and} define $\mu_*$ as
\begin{equation}
\mu_*=\frac{k_{\rm B}{{T_*}}}{3u^*r_{\rm tube}^3 \psi_*}
\ ,\end{equation}
where $k_{\rm B}$, {$T_*$}, and $u^*$ denote the Boltzmann constant,
{$T$ at $\tau=\tau_*$},
and the scaled coupling constant at the Wilson-Fisher fixed point.  
At the one loop order, $u^{*}$ equals $2\pi^{2}/9$ in the three dimensions.
Defining a dimensionless function ${\hat f}$ as
 \begin{equation}
{\hat f}({\hat\psi})={\hat\omega}^{\gamma-1}{\hat \tau}
\left( \frac{{\hat\psi}^2}{2}+\frac{{\hat\psi}^4}{12{\hat\omega}^{2\beta-1}{\hat \tau}}\right)
\ ,\label{eqn:prefmm}\end{equation}
we have {$f_-(\psi)-(\mu_-)_{\rm c}\varphi =\mu_*\psi_*T{\hat f}({\hat \psi)}/T_*$}.
The dependence of ${\hat \omega}$ on ${\hat\psi}$ and ${\hat \tau}$
is {g}iven by
\begin{equation}
{\hat \omega}={\hat \tau}+{\hat \omega}^{1-2\beta}{\hat \psi}^{2}
\ ,\label{eqn:omega2}\end{equation}  
which {comes from the self-consistent condition mentioned in Appendix \ref{sec:rlft}.}
We find that ${\hat \omega}$ as a function of ${\hat \psi}$, 
and {hence} ${\hat f}({\hat \psi})$, are even function{s.}
{We have} ${\hat f}''(0)=\hat{\tau}^{\gamma}$ {and}
\begin{equation}
f_{-}''(0)=\frac{k_{\rm B}T C_2\tau^\gamma}{3u^*\xi_0^{3}}\ .
\label{eqn:fprpr}\end{equation}
Some profiles of ${\hat \psi}^{(0)}$ are shown in Fig.~\ref{fig:lutwatvelpsi} (a) of Appendix \ref{sec:prof}.
{We define} dimensionless quantities 
 {$\delta {\hat P}$ and $\delta {\hat \mu}$ as
$\delta P/(\mu_*\psi_*)$ and $\delta\mu/\mu_*$, respectively.}
\\

\noindent
On length scales smaller than the correlation length,
correlated clusters of the order parameter are randomly convected {at equilibrium}.
 {On} larger length scales, the transport coefficient for the interdiffusion
$\Lambda$ is enhanced because the convection is averaged out, and
the viscosity $\eta_{\rm s}$ is also 
enhanced because the order parameter $\psi$ and momentum density are dynamically coupled.
{The viscosity exhibits} a weak singularity, {and hence}  
its nonuniversal background part is significant unless 
the mixture is very close to the critical point \cite{sengers,bergmold,bergmold2,iwan}. {Later},
{we use} experimental data for {h}omogeneous compositions {to} obtain
$\eta_{\rm s}$ as a function of $\tau$ and $\psi$.
This function is used even when $\psi$ is inhomogeneous. 
We introduce a dimensionless viscosity ${\hat \eta}({\hat r})\equiv \eta_0(r_{\rm tube}{\hat r})/\eta_*$, where
$\eta_*$ is defined as the value of the singular part of {$\eta_{\rm s}$} for $\tau=\tau_*$ and $\psi=0$.
{T}he flow rate
of Hagen-Poiseulle flow of a fluid, with the viscosity being 
$\eta_*$, driven by the pressure gradient $\mu_*\psi_*/L_{\rm tube}$, {is denoted by ${\cal I}_*$, which leads to} 
\begin{equation}
{\cal I}_*=\frac{ \pi r_{\rm tube}^4\mu_*\psi_* }{8\eta_* L_{\rm tube}}=\frac{\pi k_{\rm B}{T_*}r_{\rm tube}}{24 u^*\eta_*L_{\rm tube}}\ .
\end{equation}
We {define dimensionless quantities} ${\hat {\cal I}}$ and ${\hat {\cal J}}$ as
{${\cal I}/{\cal I}_*$ and ${\cal J}/(\psi_*{\cal I}_*)$, respectively.}
\\

\noindent
We extend the previous theoretical result of $\Lambda(0)$ to obtain
$\Lambda(\psi)$ for a homogeneous nonzero value of $\psi$, and then apply this extended result to a mixture
with inhomogeneous composition, as discussed in Appendix \ref{sec:traco}.
Writing $\Lambda_*$ for the value of $\Lambda$ at $\psi=0$ and $\tau=\tau_*$, we define
a dimensionless coefficient ${\hat \Lambda}({\hat r})$ as {$T\Lambda_0(r_{\rm tube}{\hat r})/(T_*\Lambda_*)$}.  
Writing $z_\psi$ for the dynamic critical exponent for the order-parameter fluctuations, we obtain
\begin{equation}
{\hat \Lambda}({\hat r})={\hat \omega}^{\nu(z_\psi-2)} \left[ {\hat f}''({\hat \psi}^{(0)}({\hat r}))\right]^{-1} 
\label{eqn:hatLamdef}\end{equation} with the aid of ${\hat f}''(0)=1$ at $\tau=\tau_*$.  We
use $z_\psi=3.067$, as mentioned in Appendix \ref{sec:traco}.
Owing to Eq.~(\ref{eqn:fprpr}), in the critical regime,
${\hat \Lambda}$ for $\psi=0$ is enhanced with approximately the same power as $\xi$ with respect to {$\tau$}.
\subsection{Formulas for the Onsager coefficients \label{sec:formulae}}
In a cylindrical tube, \textcolor{black}{we can write $v_z^{(1)}(r)$ and $j_z^{(1)}(r)$,} and
the LHS of Eq.~(\ref{eqn:st3xy+}) becomes
$\varepsilon r^{-1}\partial_r\left( r\eta_0(r) \partial_r v_z^{(1)} \right)$.
\textcolor{black}{We have $v_z^{(1)}=0$ at $r=r_{\rm tube}$
because of  the no-slip condition, and  $\partial_{r}v_{z}^{(1)}=0$ at $r=0$ 
because of the axissymmetry and smoothness of $v_z^{(1)}$.}
{Thus,} we obtain
\begin{eqnarray}&&
\varepsilon v_{z}^{(1)}(r)=
-\frac{\delta P}{L_{\rm tube} }
\int_{r}^{r_{\rm tube}}ds'\ \frac{1}{ s'{\eta_0(s')}}\int_{0}^{s'}ds\ s 
\nonumber\\ &&\qquad
-\frac{\delta \mu}{L_{\rm tube}}
\int_{r}^{r_{\rm tube}}ds'\ \frac{1}{ s'{\eta_0(s')}}\int_{0}^{s'}ds\ s \psi^{(0)}(s)
\ .\label{eqn:vzprofile2}\end{eqnarray}
Substituting
Eqs.~(\ref{eqn:jzprofile}) and (\ref{eqn:vzprofile2}) into Eqs.~(\ref{eqn:IJexpress2a}) and (\ref{eqn:IJexpress2b})
yields formulas for ${\cal L}_{ij}$.
 For convenience,
we define a dimensionless $2\times 2$ matrix ${\hat {\cal L}}$ so that
\begin{equation}
\left(\begin{array}{c} {\hat {\cal I}} \\ {\hat {\cal J}}\end{array}\right)=-{\hat {\cal L}}
\left(\begin{array}{c}  \delta {\hat P}  \\ \delta{\hat \mu}  \end{array}\right) 
\end{equation}
holds, {where $\delta {\hat P}$, $\delta {\hat \mu}$, ${\hat {\cal I}}$, and 
${\hat {\cal J}}$ are defined in the preceding subsection.}
The components of ${\hat {\cal L}}$ are not  {strictly} the Onsager coefficients because
the column vectors on both the sides above are not conjugate fluxes and forces, 
{although ${\hat {\cal L}}_{12}$
equals ${\hat {\cal L}}_{21}$.}
{We have}
\begin{equation}
{\cal L}_{11}=\frac{T{\cal I}_*}{ \mu_*\psi_*}  {\hat {\cal L}}_{11}\ ,\quad {\cal L}_{12}=\frac{T{\cal I}_*}{\mu_*}{\hat{\cal L}}_{12}\ ,
\quad  {\rm and}\quad {\cal L}_{22}=\frac{T{\cal I}_*\psi_*}{\mu_*} {\hat {\cal L}}_{22}
\ .\label{eqn:LcalL}\end{equation} 
{W}e define a functional $\Omega_{{1}}$ as
\begin{equation}
\Omega_{{1}}[g]=16 \int _0^1
dq_1\ q_1{\hat \psi}^{(0)}(q_1) \int_{q_1}^1dq_2\ \frac{1}{q_2{\hat \eta}(q_2)}\int_0^{q_2}dq_3\ q_3g(q_3)
\ ,\end{equation}
and define another functional $\Omega_0$ as the RHS above with ${\hat\psi}^{(0)}(q_1)$ being deleted.
The {formulas} for ${\cal L}_{ij}$ are given by Eq.~(\ref{eqn:LcalL}) and
\begin{eqnarray}
&&{\hat {\cal L}}_{11}=\Omega_0[1]\ ,\quad {\hat {\cal L}}_{12}=\Omega_0[{\hat \psi}^{(0)}]\ ,\quad
{\hat {\cal L}}_{21}=\Omega_{{1}}[1]\ ,\nonumber\\
&&\quad {\rm and}\quad {\hat {\cal L}}_{22}
=\Omega_{{1}}[{\hat \psi}^{(0)}]+\frac{16\pi }{9} \int_0^1d{\hat r}\ {\hat r}{\hat \Lambda}({\hat r})\ .
\label{eqn:calLcomp1}\end{eqnarray}
The coefficient {$16\pi/9$} is derived in Appendix \ref{sec:traco}.  Thanks to this,
we need not  know the value of  {$\Lambda_*$} 
in later calculations with dimensions.
The terms ${\cal L}_{22{\rm v}}$ and ${\cal L}_{22{\rm d}}$ are non-dimensionalized to give
the first and second terms on the RHS of the last entry in Eq.~(\ref{eqn:calLcomp1}), which are 
denoted by ${\hat {\cal L}}_{22{\rm v}}$ and ${\hat {\cal L}}_{22{\rm d}}$, respectively.  
Only the latter term depends on ${\hat \Lambda}$ and involves the interdiffusion among all the terms on the
\textcolor{black}{RHSs in} Eq.~(\ref{eqn:calLcomp1}), \textcolor{black}{whereas} the others depend on ${\hat \eta}$ and involve convection.
Interchanging the orders of the integrations, we have
\begin{equation}
{\hat {\cal L}}_{11}=4\int_0^1dq\ \frac{q^3}{{\hat \eta}(q)}\quad{\rm and}\quad  {\hat {\cal L}}_{12}={\hat {\cal L}}_{21}
=8\int_0^1dq_1\ q_1{\hat \psi}^{(0)}(q_1)\int_{q_1}^1dq_2\ \frac{q_2}{{\hat \eta}(q_2)}
\label{eqn:calLcomp} \ .\end{equation}
Dependence of ${\hat{\cal L}}_{11}$ on ${\hat\psi}^{(0)}$ is only via ${\hat \eta}$,
and that of ${\hat{\cal L}}_{22{\rm d}}$ is only via ${\hat \Lambda}$.  
In contras{t}, ${\hat{\cal L}}_{12}={\hat{\cal L}}_{21}$ 
and ${\hat{\cal L}}_{22{\rm v}}$ depend on ${\hat\psi}^{(0)}$ explicitly, and vanish if ${\hat\psi}^{(0)}$ vanishes,
{\it i.e.\/}, if the PA is absent, {as mentioned in the last paragraph of Section \ref{sec:sol}}.
These explicit dependences can result
from the difference in the parentheses in Eq.~(\ref{eqn:IJexpress2b}) and the same difference in Eq.~(\ref{eqn:st3xy+}).
The integrand for ${\hat{\cal L}}_{12}={\hat{\cal L}}_{21}$ in Eq.~(\ref{eqn:calLcomp}) 
contains ${\hat \psi}^{(0)}$ once because of either of the two origins, \textcolor{black}{whereas} 
the one for ${\hat{\cal L}}_{22{\rm v}}$ in Eq.~(\ref{eqn:calLcomp1}) 
 twice because of both the origins.   
This suggests that ${\hat {\cal L}}_{22{\rm v}}$ has a stronger dependence
on ${\hat \psi}^{(0)}$. 
\\

\noindent
When the pressure difference ($\delta P\ne 0$ and $\delta c_{\rm a}=0$) {is imposed, 
the mass flow rates can be calculated by
substituting} the LPEs of Eq.~(\ref{eqn:newlinph}) into Eq.~(\ref{eqn:marmbr}) and {using}
Eqs.~(\ref{eqn:kakkoP2}) and (\ref{eqn:LcalL}), and {hence}
$d {\cal M}_{n{\rm R}}/(dt)$ {equals}  $\delta P$ multiplied by
\begin{equation}
-\frac{{\cal I}_*\rho_n^{({\rm ref})} }{\mu_*\psi_*} \left( {\hat {\cal L}}_{11}+\psi_* {\bar v}_-^{({\rm ref})}{\hat {\cal L}}_{12}\right)
\mp\frac{{\cal I}_*}{2\mu_*} \left( {\hat {\cal L}}_{21}
+ \psi_* {\bar v}_-^{({\rm ref})} {\hat {\cal L}}_{22}\right)
\ ,\label{eqn:mfr1}\end{equation}
where the double sign implies the same as in Eq.~(\ref{eqn:marmbr}).
\textcolor{black}{Notably, the interdiffusion is accompanied with the Hagen-Poiseulle flow
in the absence of PA $(h=0$) unless ${\bar v}_-^{({\rm ref})}$ vanishes.}
When {t}he mass-fraction difference ($\delta P=0$ and $\delta c_{\rm a}\ne 0$) {is imposed},
$d {\cal M}_{n{\rm R}}/(dt)$ {equals}  $\left(\delta \mu\right)_P$ multiplied by
\begin{equation}
-\frac{{\cal I}_*\rho_n^{({\rm ref})} }{\mu_*} {\hat {\cal L}}_{12}\mp\frac{{\cal I}_*\psi_*}{2\mu_*} {{\hat {\cal L}}_{22}}
\ .\label{eqn:mfr2}\end{equation}
\section{Numerical results\label{sec:num}}
{Our} numerical results are obtained 
with the aid of the software Mathematica (Wolfram Research).
The tube radius is put equal to $r_{\rm tube}=0.1\ \mu$m. 
In Figs.~\ref{fig:lutwatvelpsi}  (b) and (c) of Appendix \ref{sec:prof},
we show some flow profiles, which are consistent with
our coarse-grained description.
{We abbreviate 2,6-lutidine and water to LW, and nitroethane and 3-methylpentane to NEMP.}
\subsection{Onsager coefficients for a mixture of LW \label{sec:LW}}
\begin{figure}
\includegraphics[width=12cm]{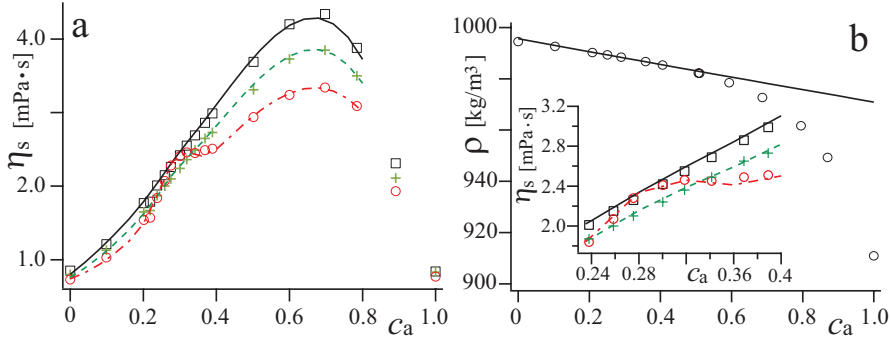}
\caption{Experimental data for a mixture of LW; $c_{\rm a}$ denotes the mass fraction of 2,6-lutidine.
(a) Data of viscosity \textcolor{black}{$\eta_{\rm s}$}
at $\tau=2.25\times 10^{-2}$, $1.27\times10^{-2}$ and $7.82\times 10^{-5}$ 
in Ref.~\cite{stein} are replotted with the black square, green cross, and red circle, respectively.  
Curve fittin{g} for $c_{\rm a}<0.8$
yield the black solid, green dashed, and red dash-dot curves for the respective values of $\tau$.
These values of $\tau$ are different from the values of $\tau$ used in the following figures of our numerical
results for a mixture of LW.
(b) The circle represents experimental data from Table III of Ref.~\cite{jaya} for the relationship between
the total mass density $\rho$ and the mass fraction $c_{\rm a}$ at $306.65\ $K.
The solid curve is obtained by {curve fitting}
to the data for $0.2<c_{\rm a}<0.41$.
The inset shows a magnified view of a part of (a).
}
\label{fig:viscosity}
\end{figure}

\begin{figure}
\includegraphics[width=12cm]{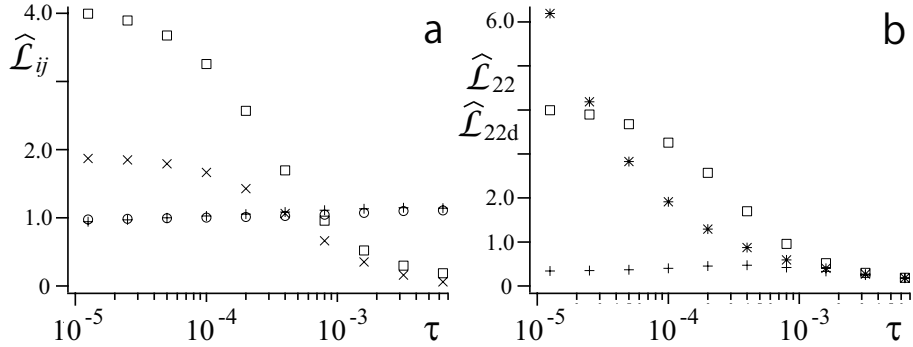}
\caption{Plots of the dimensionless Onsager coefficients against the reduced temperature $\tau$ 
for a mixture of LW. 
(a) The symbols $\circ$, $\times$, and $\square$
represent ${\hat{\cal L}}_{11}$, ${\hat{\cal L}}_{12}$, and ${\hat{\cal L}}_{22}$, respectively, calculated by setting
the surface field $h$ equal to $0.1\ $cm$^3/$s$^2$.
The symbol $+$, representing ${\hat{\cal L}}_{11}$ for $h=0$, is obtained by replacing the dimensionless viscosity 
${\hat \eta}({\hat r})$ with its value at the critical composition irrespective of ${\hat r}$.
(b) The symbol $\square$ represents ${\hat{\cal L}}_{22}$, and is a replot of the same symbol in (a). The symbol $+$
represents ${\hat {\cal L}}_{22{\rm d}}$, which is a part of ${\hat{\cal L}}_{22}$, for $h=0.1\ $cm$^3/$s$^2$, whereas
the symbol $*$ represents ${\hat{\cal L}}_{22{\rm d}}$ for $h=0$.  
}
\label{fig:lutwat}
\end{figure}

We consider a mixture of LW {n}ear the  
lower critical consolute point,  taking 2,6-lutidine 
to be the component a.
The value of $\xi_0=1.98\times 10^{-1}\ $nm in Ref.~\cite{mirz} gives
$\tau_*=5.12\times 10^{-5}$; we use ${T}_{\rm c}=307\ $K and $c_{\rm ac}=0.290$.  The estimate
of $C_2$ in {Appendix \ref{sec:rlft}} gives 
$\psi_*=4.70\times 10\ $kg$/$m$^3$, which leads to $\mu_*=1.37\times 10^{-2}\ $m$^2/$s$^2$.
{I}n Appendix \ref{sec:traco}, we obtain 
the viscosity {a}s a function of the reduced temperature $\tau$
and the order parameter $\psi$  {and} find $\eta_*=2.44\times 10^{-3}\ $Pa$\cdot$s
{by} using the experimental data of Refs.~\cite{gratt,stein}, 
part of which are replotted in Fig.~\ref{fig:viscosity}(a).
In this figure, {the curves represent the function.}  {Each of the} 
black solid and  green dashed curves appear{s} to be composed of 
the background part of $\eta_{\rm s}$ alone,
whereas a small peak due to the singular part {appears on}
the red dash-dot curve. {This} curve has a nearly flat portion immediately to the right of the peak,
which is clearly shown in the inset of  Fig.~\ref{fig:viscosity}(b).  
 The background part {is} asymmetric about the vertical line at $c_{\rm a}=c_{\rm ac}$.
{I}n the homogeneous phase near the critical temperature, how $\rho$ depends on $c_{\rm a}$ is measured 
in Ref.~\cite{jaya}.  
{With} the partial volumes, ${\bar v}_{\rm a}$ and ${\bar v}_{\rm b}$, assumed to be constants, 
$\rho$ is {a}  function of $c_{\rm a}$ {via}
Eq.~(\ref{eqn:rhovv}). 
The {curve-fit} of the function to the data in Fig.~\ref{fig:viscosity}(b)
gives ${\bar v}_{\rm a}=1.03\times 10^{-3}\ $m$^3/$kg and  ${\bar v}_{\rm b}=1.00\times 10^{-3}\ $m$^3/$kg,
which are respectively regarded as ${\bar v}_{\rm a}^{({\rm ref})}$ and ${\bar v}_{\rm b}^{({\rm ref})}$.
These values yield $\rho_{\rm c}=9.88
\times 10^2\ $kg$/$m$^3$ and $\varphi_{\rm c}=-4.15\times 10^2\ $kg$/$m$^3$.
In the following calculations, 
{$\left|c_{\rm a}-c_{\rm ac}\right|$} in the tube at equilibrium is smaller than $0.15$.
According to the discussion in Section 6 of Ref.~\cite{yabufuji}, the magnitude of the surface field $h$
should be smaller than approximately $10\ $cm$^3/$s$^2$ when
2,6-lutidine adsorbs onto the solid surface.\\
\noindent
Assuming $h=0.1\ $cm$^3/$s$^2$, we use Eqs.~(\ref{eqn:calLcomp1}) and (\ref{eqn:calLcomp}) to obtain
the results shown
in Fig.~\ref{fig:lutwat}(a), \textcolor{black}{where} ${\hat {\cal L}}_{12}$ is close to zero at $\tau=6.4\times10^{-3}$ and
 increases eminently as $\tau$ decreases.  Then,
${\hat {\cal L}}_{22}$ increases more eminently.  
From Fig.~\ref{fig:lutwat}(b), we find that
${\hat {\cal L}}_{22{\rm v}}={\hat {\cal L}}_{22}-{\hat {\cal L}}_{22{\rm d}}$ are close to zero  at $\tau=6.4\times10^{-3}$,
and that the increase of ${\hat {\cal L}}_{22}$ with decreasing $\tau$ is mainly caused by that of ${\cal L}_{22{\rm v}}$.
These findings are expected from the statements 
{i}n the penultimate paragraph of Section \ref{sec:formulae}; 
{some plots of ${\hat \psi}^{(0)}$ are given in Fig.~\ref{fig:lutwatvelpsi}(a).}
In the absence of PA, noting that
${\hat \Lambda}$ becomes homogeneously equal to the value at the
critical composition {in Eq.~(\ref{eqn:calLcomp1})}, we calculate ${\hat {\cal L}}_{22{\rm d}}$ and plot the results 
in Fig.~\ref{fig:lutwat}(b).  They show a significant increase with decreasing $\tau$, 
which is caused by the critical enhancement of $\Lambda$. 
{In the presence of PA,} decrease in $\tau$ tends to cause the critical enhancement and
tends to shift the composition away from the critical one (Fig.~\ref{fig:lutwatvelpsi}(a)).
As a result of these conflicting tendencies,
the values  shown by the cross in Fig.~\ref{fig:lutwat}(b) are approximately independent on $\tau$.
\\

\begin{figure}
\includegraphics[width=12cm]{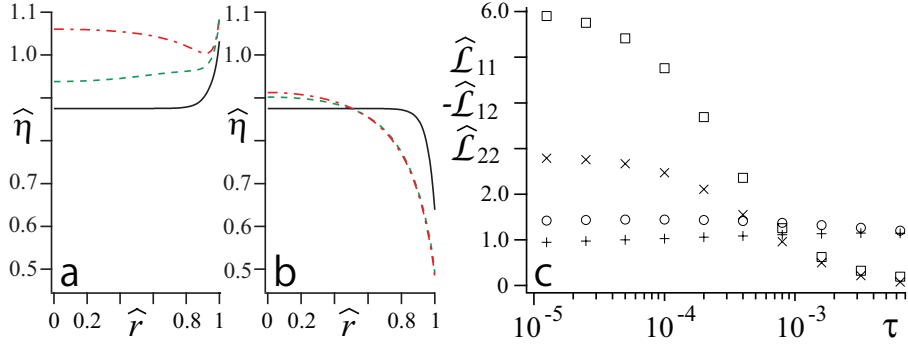}
\caption{We use $h=0.1\ $cm$^3/$s$^2$ in (a) and
$h=-0.1\ $cm$^3/$s$^2$ in (b) to plot the dimensionless viscosity
${\hat \eta}({\hat r})$ for a mixture of LW against ${\hat r}$. 
The black solid, green dashed, red dash-dot curves represent the results
for $\tau=6.4\times 10^{-3}$,
$4.0\times 10^{-4}$, and $1.25\times 10^{-5}$, respectively.  For these values of $\tau$,
${\hat\eta}$ takes $0.875, 0.922$ and $1.06$ at the critical composition, respectively.
We use $h=-0.1\ $cm$^3/$s$^2$ to plot
${\hat{\cal L}}_{11}$ $(\circ)$, $-{\hat{\cal L}}_{12}$ $(\times)$, and ${\hat{\cal L}}_{22}$ $(\square)$
against $\tau$ in (c).
For comparison, the same results as shown by the cross in Fig.~\ref{fig:lutwat}(a) are here replotted with the same symbol.
}
\label{fig:lutwatvisc}
\end{figure}

\noindent
When the sign of $h$ is changed, that of ${\hat \psi}^{(0)}$ changes, {as mentioned below 
Eq.~(\ref{eqn:stationary}).  This is reasonable because the changes can be made by
renaming the components in the opposite way.}
For the two smallest values of $\tau$, $\eta_{\rm s}$ changes less
 in Fig.~\ref{fig:lutwatvisc}(a) for $h>0$ than in Fig.~\ref{fig:lutwatvisc}(b) for $h<0$
because $c_{\rm a}$ in the tube is larger than $c_{\rm a c}$ for $h>0$ and
  {the red dash-dot curve has the nearly flat portion}
in Fig.~\ref{fig:viscosity}(a).
Thus, the results for $h>0$ $(\circ)$
and the ones for $h=0$ $(+)$ approximately agree in Fig.~\ref{fig:lutwat}(a), and not in Fig.~\ref{fig:lutwatvisc}(c).
In this figure,
${\hat {\cal L}}_{12}$ becomes negative because of ${\hat \psi}^{(0)}<0$ in Eq.~(\ref{eqn:calLcomp}), whereas
 ${\hat {\cal L}}_{22}$ is positive because of Eqs.~(\ref{eqn:hatLamdef}) and (\ref{eqn:calLcomp1}).
In Figs.~\ref{fig:lutwat}(a) and \ref{fig:lutwatvisc}(c),
$|{\hat {\cal L}}_{12}|$ and ${\hat {\cal L}}_{22}$ increases as $\tau$ decreases in the presence of PA,
because $|{\hat \psi}^{(0)}({\v r})|$ {in} $\Omega_{{1}}[1]$ and $\Omega_{{1}}[{\hat \psi}^{(0)}]$
then increases {(Fig.~\ref{fig:lutwatvelpsi}(a))}.
When the sign of $h(\ne 0)$ is changed at a value of $\tau$, the value of $\left|{\hat {\cal L}}_{ij}\right|$ 
does not remain the same 
because of the asymmetry of the background part of $\eta_{\rm s}$ in Fig.~\ref{fig:viscosity}(a).
Then, ${\hat{\cal L}}_{22{\rm d}}$ remains unchanged owing to Eqs.~(\ref{eqn:hatLamdef}) and 
(\ref{eqn:calLcomp1}).
When we set $\delta P=0$ and $\delta c_{\rm a}<0$, which gives $\left(\delta\mu\right)_P=\delta\mu<0$ 
owing to Eq.~(\ref{eqn:kakkoP2}), 
the total mass flow enters into the right (left) reservoir through the tube with $h>0$ ($h<0$)
because of Eq.~(\ref{eqn:mfr2}) and
${\hat {\cal L}}_{12}>0$ (${\hat {\cal L}}_{12}<0$).
This flow direction is reasonable, considering that it \textcolor{black}{would} relax the mass-fraction difference between the
reservoirs because the component a is \textcolor{black}{concentrated in (depleted from) the tube with $h>0$ ($h<0$)}.
\subsection{Onsager coefficients for a mixture of NEMP \label{sec:NEMP}}
\begin{figure}
\includegraphics[width=12cm]{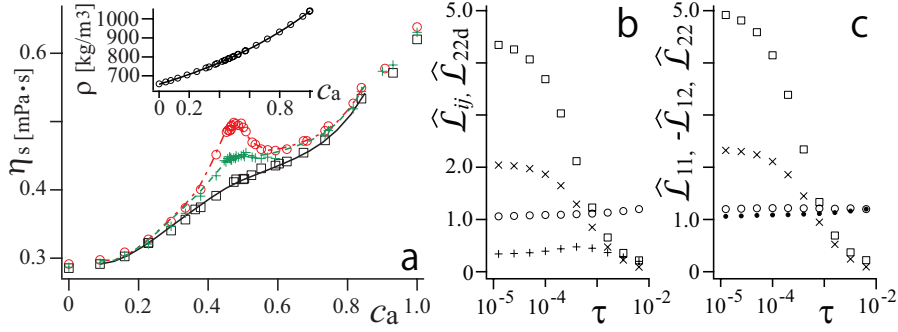}
\caption{(a) Experimental data for a mixture of NEMP; $c_{\rm a}$ denotes the mass fraction of nitroethane.
We shift the data of viscosity at $\tau=1.47\times 10^{-4}$, $1.82\times 10^{-3}$, and $1.19\times 10^{-2}$ \textcolor{black}{of} Ref.~\cite{leis}, 
as \textcolor{black}{mentioned} in Appendix \ref{sec:traco}, to plot the results with red circles, green crosses, and black squares, respectively.  
Curve{-fits} for $0.08<c_{\rm a}<0.85$
yield the red dash-dot, green dashed, and  black solid curves for the respective values of $\tau$.
In the inset, the circle represents experimental data from Ref.~\cite{reeder} for the relationship between $\rho$ and $c_{\rm a}$ at $300.153\ $K.
The solid curve is obtained by {curve fitting to} the data for $0.3<c_{\rm a}<0.6$, {as in Fig.~\ref{fig:viscosity}(b)}.
(b, c) \textcolor{black}{Calculation results of} the components of ${\hat{\cal L}}$ \textcolor{black}{are plotted} against $\tau$ 
for a mixture of NEMP.
The symbols $\circ$, $\times$, $\square$, and $+$ in (b)
represent ${\hat{\cal L}}_{11}$, ${\hat{\cal L}}_{12}$, ${\hat{\cal L}}_{22}$, and ${\hat{\cal L}}_{22{\rm d}}$, respectively, 
for  $h=0.1\ $cm$^3/$s$^2$. 
The symbols $\circ$, $\times$, and $\square$ in (c)
represent
${\hat{\cal L}}_{11}$, $-{\hat{\cal L}}_{12}$, and ${\hat{\cal L}}_{22}$,
respectively, 
for  $h=-0.1\ $cm$^3/$s$^2$. 
For comparison, the same results as shown by the circle in (b) are replotted with the solid circle in (c).  }
\label{fig:viscoMPNE}
\end{figure}

We consider a mixture of NEMP {n}ear the  
upper critical consolute point 
(${T}_{\rm c}=300\ $K and $c_{\rm ac}=0.466$), taking nitroethane
to be the component a.  The ratio $\rho_{\rm bc}/\rho_{\rm ac}$ is closer to unity
than the ratio for a mixture of LW.
We have $\xi_0=2.30\times 10^{-1}\ $nm \cite{iwan} to obtain
$\tau_*=6.49\times 10^{-5}$.  The estimate of $C_2$ 
in {Appendix \ref{sec:rlft}} gives
 $\psi_*=4.19\times 10\ $kg$/$m$^3$, 
which leads {to} $\mu_*=1.50\times 10^{-2}\ $m$^2/$s$^2$.  
{W}e use the experimental data of Refs.~\cite{leis,iwan} 
to obtain $\eta_*=5.10\times 10^{-4}\ $Pa$\cdot$s and dependence of
$\eta_{\rm s}$ on $\tau$ and $\psi$ in Appendix \ref{sec:traco}.
{The dependence is shown in Fig.~\ref{fig:viscoMPNE}(a), where
the} range of values of $\eta_{\rm s}$ at a value of $\tau$  is 
narrower than the one at any value of $\tau$ in Fig.~\ref{fig:viscosity}(a).  The fitted curve in 
the former figure shifts upward as $\tau$ decreases,
unlike the fitted curve  in the latter figure, which shows data
near the lower consolute point. The peak 
due to the singular part is clearer in the former figure.
{As} in
Fig.~\ref{fig:viscosity}(b), we plot $\rho$ against $c_{\rm a}$
in the inset of Fig.~\ref{fig:viscoMPNE}(a){,}  
where data from Ref.~\cite{reeder} {and the fitted curve are shown. 
The curve-fit gives}
${\bar v}_{\rm a}^{({\rm ref})}=9.60\times 10^{-4}\ $m$^3/$kg and  ${\bar v}_{\rm b}^{({\rm ref})}=1.53\times 10^{-3}\ $m$^3/$kg, 
which yield $\rho_{\rm c}=7.92\times 10^2\ $kg$/$m$^3$ and
$\varphi_{\rm c}=-5.46\times 10\ $kg$/$m$^3$. 
 In the following calculations, {$\left|c_{\rm a}-c_{\rm ac}\right|$} in the tube at equilibrium is smaller than $0.17$. \\


\noindent
The components of ${\hat{\cal L}}$ are calculated 
using Eqs.~(\ref{eqn:calLcomp1}) and (\ref{eqn:calLcomp}).
As $\tau$ decreases, ${\hat {\cal L}}_{12}={\hat{\cal L}}_{21}$ 
becomes positively larger in Fig.~\ref{fig:viscoMPNE}(b)  for $h>0$ and negatively larger Fig.~\ref{fig:viscoMPNE}(c) for $h<0$.
Then, \textcolor{black}{in either of the figures}, ${\hat{\cal L}}_{22}$ \textcolor{black}{more} 
eminently increases, \textcolor{black}{whereas} ${\hat {\cal L}}_{11}$ and
 ${\hat {\cal L}}_{22{\rm d}}$ exhibit much less distinct changes. 
At $\tau=1.25\times 10^{-5}$ for $h=0$, ${\hat {\cal L}}_{22{\rm d}}$ equals $7.08$,
which is much larger than \textcolor{black}{the corresponding value for $\left| h \right|=0.1\ $cm$^3/$s$^2$, shown by}
the cross on the extreme left in Fig.~\ref{fig:viscoMPNE}(b).
Thus, the eminent increase of ${\hat {\cal L}}_{22}$ shown in each of  Figs.~\ref{fig:viscoMPNE}(b) and (c)
is not caused by the critical enhancement of $\Lambda$.
These properties are qualitatively the same as shown 
in Figs.~\ref{fig:lutwat}(a) and \ref{fig:lutwatvisc}(c).
{The value shown by the circle on the extreme left  in 
Fig.~\ref{fig:viscoMPNE}(b) changes to  $0.935$ when $h$ vanishes.
This change is larger in magnitude than the corresponding change} in Fig.~\ref{fig:lutwat}(a) for a mixture of LW,
{owing to} the clear peak of $\eta_{\rm s}$ in Fig.~\ref{fig:viscoMPNE}(a).
\subsection{Mass flow rates \label{sec:massflow}}
We have $\psi_* {\bar v}_-^{({\rm ref})}=6.00\times 10^{-4}$ in a mixture of LW
and $-1.18 \times 10^{-2}$ in a mixture of NEMP.  In the presence of PA,
$\left|{\hat {\cal L}}_{12}\right|$
is larger than approximately one thirds of ${\hat {\cal L}}_{22}$ 
{i}n Figs.~\ref{fig:lutwat}(a), \ref{fig:lutwatvisc}(c), \ref{fig:viscoMPNE}(b), and \ref{fig:viscoMPNE}(c).
Thus, for these results under $h\ne 0$, the second terms in the two pairs of parentheses
in Eq.~(\ref{eqn:mfr1}) are negligible and
Eq.~(\ref{eqn:mfr1}) can be approximated to be
\begin{equation}
-\frac{{\cal I}_*\rho_n^{({\rm ref})} }{\mu_*\psi_*} {\hat {\cal L}}_{11}
\mp\frac{{\cal I}_*}{2\mu_*}  {\hat {\cal L}}_{21}
\ .\label{eqn:mfr1a}\end{equation}
Here, owing to $\psi_* \ll \rho_n^{({\rm ref})}=\rho_{n {\rm c}}$, the first term is predominant 
for the values of ${\hat {\cal L}}_{11}$ and ${\hat {\cal L}}_{12}$ in the figures mentioned above.
For these values, the first term is predominant in Eq.~(\ref{eqn:mfr2}) unless $h$ vanishes, 
although ${\hat {\cal L}}_{22}$ is larger than $\left|{\hat {\cal L}}_{12}\right|$ as far as examined.
We define a dimensionless mass flow rate of the component $n$ as
\begin{equation}
\frac{d{\hat {\cal M}}_{n{\rm R}}}{d{\hat t}} \equiv 
{\frac{1}{\rho_{\rm c}{\cal I}_*}}
\frac{d{\cal M}_{n{\rm R}}}{dt}  
\ .\label{eqn:dimmass}\end{equation}
{A} dimensionless total mass flow rate, denoted by $d{\hat {\cal M}}_{\rm R}/(d{\hat t})$,  
is defined as the sum of Eq.~(\ref{eqn:dimmass}) over $n=$ a and b, {and is equal to ${\hat {\cal I}}$.}
Below, in flow driven by the pressure difference, we calculate
 $c_{\rm aR}^{({\rm flux})}$, which is defined  as the ratio of
Eq.~(\ref{eqn:dimmass}) with $n=$a to $d{\hat {\cal M}}_{\rm R}/(d{\hat t})$. This ratio 
can be interpreted as
the mass fraction of the component a in a mixture flowing into the right reservoir per unit time
when the mass flow rates {of the components} share the same sign,  {becoming} equal to $c_{\rm ac}$ when
the second term of Eq.~(\ref{eqn:mfr1}) vanishes. \\

\begin{figure}
\includegraphics[width=12cm]{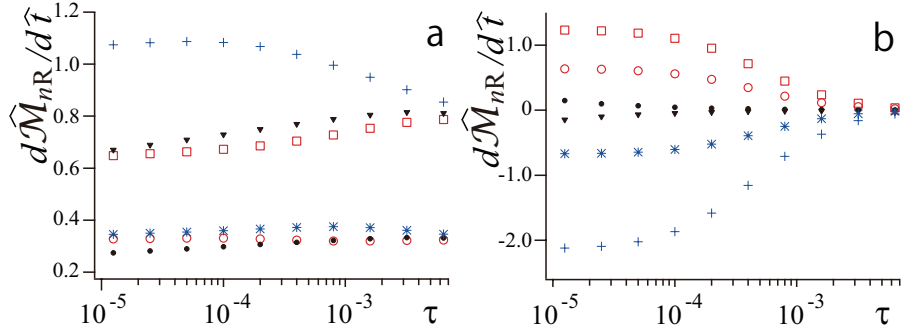}
\caption{Plots of the dimensionless mass flow rates against $\tau$ for a mixture of LW;
$d{\hat {\cal M}}_{\rm aR}/(dt)$ is plotted with red open circles, solid circles, and blue asterisks, whereas
$d{\hat {\cal M}}_{\rm bR}/(dt)$ with red squares, solid triangles, and blue crosses. 
The values of open circles and squares are obtained for $h=0.1\ $cm$^3/$s$^2$, 
those of solid circles and triangles for $h=0$,
and those of asterisks and crosses for   $h=-0.1\ $cm$^3/$s$^2$.
We {set $(\delta P, (\delta  \mu)_P)$ to be $(-\mu_*\psi_*, 0)$}
in (a), and {$(0,-\mu_*)$ in (b).}
}
\label{fig:lutwatmnr}
\end{figure}

\noindent
For a mixture of LW, 
the dimensionless mass flow rates under {$(\delta P, (\delta  \mu)_P)=(-\mu_*\psi_*, 0)$}
are calculated by using Eq.~(\ref{eqn:mfr1}) and plotted 
in Fig.~\ref{fig:lutwatmnr}(a).
\textcolor{black}{For $h>0$}, the mass flow rate of the component a
is less dependent on $\tau$ than that of {b}, \textcolor{black}{as shown by open circles and squares.}
This is consistent wit{h} Fig.~\ref{fig:lutwatvisc}(a), {where} the curves {o}f ${\hat \eta}$ 
are close to each other in the neighborhood of the tube wall {attracting} the component a. 
For these mass flow rates, $c_{\rm aR}^{({\rm flux})}$ increases from $0.291$ to $0.336$ 
as $\tau$ decreases from $6.4\times 10^{-3}$ to $1.25\times 10^{-5}$.
The deviation from $c_{\rm a c}$ is mainly caused by the second term 
in Eq.~(\ref{eqn:mfr1a}), and $c_{\rm aR}^{({\rm flux})}>c_{\rm ac}$
can be realized intuitively because $h>0$ means the absorption of \textcolor{black}{the} component a in the tube.
\textcolor{black}{For $h=0$, noting ${\hat{\cal L}}_{12}={\hat{\cal L}}_{21}={\hat{\cal L}}_{22{\rm v}}=0$,  
we substitute} the values of the crosses in Fig.~\ref{fig:lutwat}(a)
and those of the asterisks in  Fig.~\ref{fig:lutwat}(b) into \textcolor{black}{${\hat {\cal L}}_{11}$ and ${\hat {\cal L}}_{22}$ of}
Eq.~(\ref{eqn:mfr1}), \textcolor{black}{respectively, to} obtain the results shown by the solid circle and triangle in Fig.~\ref{fig:lutwatmnr}(a).
They {d}ecrease {as $\tau$ decreases}{, in particular} for small $\tau$,
\textcolor{black}{which is attributed to
the change of the first term of Eq.~(\ref{eqn:mfr1a}) due to} the weak critical enhancement of $\eta_{\rm s}$.
For them, because the second term of Eq.~(\ref{eqn:mfr1a}) vanishes, $c_{\rm aR}^{({\rm flux})}$ remains close to $c_{\rm a c}$,
ranging  from $c_{\rm ac}=0.2900$ to $0.2901$.
For $h<0$, \textcolor{black}{the results are }shown by the asterisk and cross in Fig.~\ref{fig:lutwatmnr}(a),
\textcolor{black}{and} $c_{\rm aR}^{({\rm flux})}$ decreases from $0.289$ to  $0.243$ 
as $\tau$ decreases from $6.4\times 10^{-3}$ to $1.25\times 10^{-5}$.
{They show $c_{\rm aR}^{({\rm flux})}<c_{\rm ac}$} 
 because $h<0$ means {the depletion of} the component a from the tube.
The value shown by the cross noticeably increases as $\tau$ decreases from $\tau=6.4\times 10^{-3}$ to
$\tau=4.0\times 10^{-4}$ in Fig.~\ref{fig:lutwatmnr}(a), \textcolor{black}{which} is consistent with \textcolor{black}{the results of ${\hat \eta}$
 in Fig.~\ref{fig:lutwatvisc}(b).   In this figure,} the dashed and dash-dot curves 
are distinctly below the solid curve
in the neighborhood of the tube wall {attracting} the component b.\\

\noindent
For a mixture of LW, the dimensionless mass flow rates 
under {$(\delta  P, (\delta \mu)_P)=(0, -\mu_*)$}  are calculated 
by using Eq.~(\ref{eqn:mfr2}) and plotted 
in Fig.~\ref{fig:lutwatmnr}(b).  
The results for $h>0$, shown by the open circle and square,
increase distinctly as $\tau$ decreases.  
At $\tau=1.25\times 10^{-5}$, the ratio of the first term of Eq.~(\ref{eqn:mfr2}) to its total
is $0.851$ for the component a and the ratio for b is $1.07$.  These values are rather close to unity, indicating that 
the first term is dominant.
In the absence of \textcolor{black}{PA}, we have 
${\hat {\cal L}}_{12}={\hat {\cal L}}_{22{\rm v}}=0$, and
the total mass flow rate vanishes.  This can be
read from the results shown by the solid circle and triangle in Fig.~\ref{fig:lutwatmnr}(b).  
They increase in magnitude as $\tau$ decreases, which represents the critical enhancement of
$\Lambda$.  
The results for $h<0$, shown by
the asterisk and cross, become negatively larger as
$\tau$ decreases in Fig.~\ref{fig:lutwatmnr}(b). 
At $\tau=1.25\times 10^{-5}$, the ratio of the first term of Eq.~(\ref{eqn:mfr2}) to its total
is $1.21$ for the component a and the ratio for b is $0.937$.  These values indicate the predominance of the first term.
Thus, the total mass flow rate is positively (negatively) larger for $h>0$ ($h<0)$ as $\tau$ decreases,
which is expected from the change of ${\hat {\cal L}}_{12}$ in Fig.~\ref{fig:lutwat}(a) (Fig.~\ref{fig:lutwatvisc}(c))
and implies that diffusioosmosis is caused by the PA. 
\\

\begin{figure}
\includegraphics[width=12cm]{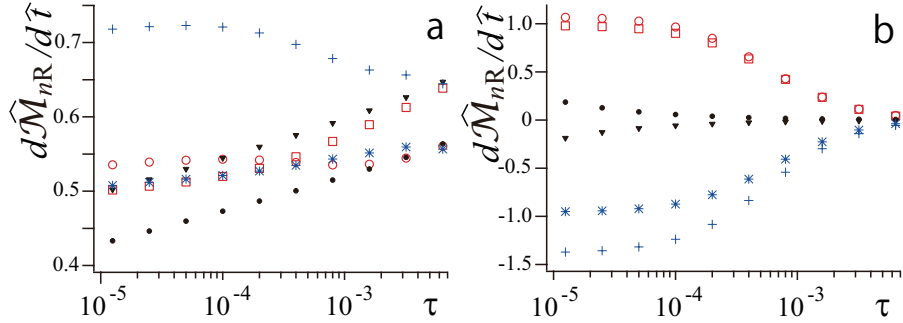}
\caption{Plots of \textcolor{black}{$d{\hat {\cal M}}_{n{\rm R}}/(d{\hat t})$ against $\tau$ for a mixture of NEMP,
corresponding with Fig.~\ref{fig:lutwatmnr}.
{T}he symbols in (a) and (b) have the same meanings as the ones in (a) and (b) in Fig.~\ref{fig:lutwatmnr}, respectively.}
 }
\label{fig:mpnemnr}
\end{figure}

\noindent
Qualitatively, the dimensionless mass flow rates for a mixture of NEMP
share many properties with those for a mixture of LW.  For {that} mixture,
the dimensionless mass flow rates under {$(\delta P, (\delta \mu)_P)=(-\mu_*\psi_*, 0)$ and
$(\delta P, (\delta \mu)_P)=(0, -\mu_*)$
are plotted in Figs.~\ref{fig:mpnemnr}(a) and (b), respectively.  In the former},  
for {the results shown by the open circle and square
($h>0$),} $c_{\rm aR}^{({\rm flux})}$ increases from $0.467$ to $0.516$ as $\tau$ decreases from $6.4\times 10^{-3}$ 
to $1.25\times 10^{-5}$.
The results for $h=0$, shown by the solid circle and triangle in Fig.~\ref{fig:mpnemnr}(a), decrease {as $\tau$ decreases}.
For these values, $c_{\rm aR}^{({\rm flux})}$ remains close to $c_{\rm ac}$, ranging from $0.463$ to $c_{\rm ac}=0.466$.
For $h<0$, the results are shown by the asterisk and cross in Fig.~\ref{fig:mpnemnr}(a), \textcolor{black}{and}
$c_{\rm aR}^{({\rm flux})}$ decreases from $0.463$ to $0.414$ as 
$\tau$ decreases from $6.4\times 10^{-3}$ to $1.25\times 10^{-5}$.  
{In Fig.~\ref{fig:mpnemnr}(b), the} results for $h>0$, shown by the open circle and square,
increase eminently as $\tau$ decreases.  
At $\tau=1.25\times 10^{-5}$, the ratio of the first term of Eq.~(\ref{eqn:mfr2}) to its total
is $0.893$ for the component a and the ratio for b is $1.12$.
The magnitudes of the results for $h=0$, shown by the solid circle and
triangle in  Fig.~\ref{fig:mpnemnr}(b),
 increase as
$\tau$ decreases, and the total mass flow rate vanishes.
The results for $h<0$, shown by the asterisk and cross in Fig.~\ref{fig:mpnemnr}(b),
become negatively larger as $\tau$ decreases.
The ratio of the first term of Eq.~(\ref{eqn:mfr2}) to its total
is $1.14$ for the component a and the ratio for b is $0.906$.  
{T}he magnitude of the total mass rate increases with decreasing $\tau$ for $h\ne 0$, which implies that
diffusioosmosis is caused by the PA. 
\subsection{Diffusioosmosis\label{sec:diffusio}}
\begin{figure}
\includegraphics[width=12cm]{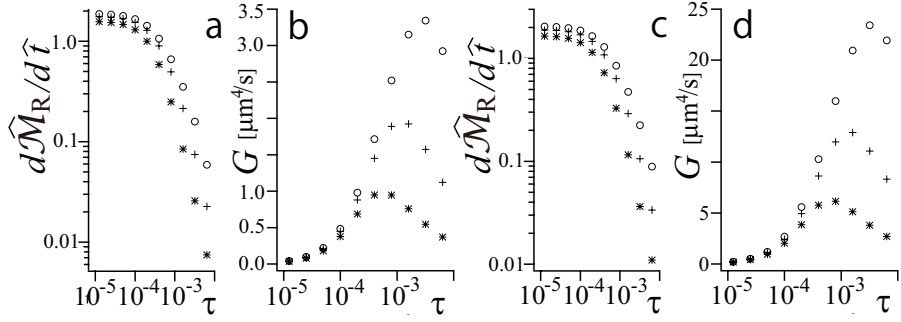}
\caption{{W}e plot the dimensionless total mass flow rate $d{\hat {\cal M}}_{{\rm R}}/(d{\hat t})$  for 
${(\delta P, (\delta \mu)_P)
 = (0, -\mu_*)}$ in (a) and
the conductance $G$ in (b) {a}gainst $\tau$ for a mixture of LW.
Those for a mixture of NEMP are plotted in (c) and (d), respectively.
Circles, crosses, and asterisks represent the results obtained by setting the surface field $h$
equal to $0.1$, $0.032$, 
and $0.01\ $cm$^3/$s$^2$, respectively. 
The result at a value of $\tau$ for $h=0.1\ $cm$^3/$s$^2$ in (a)    
is the sum of the results shown by the open circle and square at the value of $\tau$
in Fig.~\ref{fig:lutwatmnr}(b).
The same relationship holds between the results in (c) and Fig.~\ref{fig:mpnemnr}(b).}
\label{fig:osmcoeff}
\end{figure}

The first two terms on the RHS of Eq.~(\ref{eqn:stok}) can be rewritten as
$-\nabla \left(p+\varphi_{\rm c}\mu\right)-\psi\nabla\mu$ with the aid of Eq.~(\ref{eqn:Piosmgen1}).  
The sum $p+\varphi_{\rm c}\mu$ represents a scalar pressure, which
keeps the incompressibility at the order of $\varepsilon$.  In the presence of PA,
the term $-\psi\nabla\mu$ {becomes nonzero to cause diffusioosmosis} 
when a gradient of $\mu$ is imposed.  {Being significant under the mass-fraction difference, this mechanism
also works under the pressure difference to yield a slight contribution, as shown by the second term in the first parenthesis of
Eq.~(\ref{eqn:mfr1}).}
{I}n the linear regime considered in the present study,
{we} define $G$ so that
\begin{equation}
\frac{1}{\rho_{\rm c}} \frac{d {\cal M}_{\rm R}}{dt} =-G \frac{\delta c_{\rm a}}{c_{\rm ac}L_{\rm tube}}\ \label{eqn:flrate}
\end{equation} holds. {Here,} 
the {LHS} equals {the flow rate ${\cal I}$ caused by}  
the mass-fraction difference $\delta c_{\rm a}$, with $\delta P$ setting equal to zero, 
whereas the fraction on the RHS is th{e} {composition}
gradient normalized by $c_{\rm ac}$.
Thus, $G$ represents the conductance in diffusioosmosis {a}nd {equals the product of
Eq.~(\ref{eqn:fromdiffo}) and ${\cal I}_* c_{\rm a c}L_{\rm tube} {\hat L}_{12}/\mu_*$, where 
${\hat L}_{12}$ equals
$d{\hat {\cal M}}_{\rm R}/(d{\hat t})\equiv {\hat {\cal I}}$ 
under $( \delta P,   (\delta \mu)_P  )=(0,  -\mu_* )$. 
The signs of $G$ and $h$ are the same because of the discussion on Figs.~\ref{fig:lutwatmnr}(b) and \ref{fig:mpnemnr}(b).}
As shown in Fig.~\ref{fig:osmcoeff}(a) for a mixture of LW and Fig.~\ref{fig:osmcoeff}(c)
for a mixture of NEMP, the dimensionless total mass flow rate {for $h>0$} increases as $\tau$ decreases,
and appears to reach a plateau for small $\tau$.  
The increase is due to that of the adsorption layer's thickness, which is caused by 
that of the osmotic susceptibility. 
{The plateau would come from the finite size of the tube's cross section.} 
{The increase of the susceptibility also reduces  
$\left|(\delta\mu)_P\right|$} under a given value of $\delta c_{\rm a}$, {and hence}
$G$ decreases for small $\tau$ {i}n Figs.~\ref{fig:osmcoeff}(b) and (d) {for $h>0$}.
\textcolor{black}{Owing to the plateau mentioned above, Eq.~(\ref{eqn:fromdiffo}), and Eq.~(\ref{eqn:fprpr}),
$G$ is approximately proportional} to $\tau^\gamma$ for small $\tau$.
For each value of $h$, $G$ has a peak in these figures.  
The value of $\tau$ at the peak decreases {as $h$ decreases}, because then {occurrence of} the plateau 
is limited to smaller values of $\tau$ in Figs.~\ref{fig:osmcoeff}(a) and (c).
For a value of $h$, the maximum of $G$ in the latter figure are larger than the {one} in the former 
because $\eta_{\rm s}$ takes smaller values in a mixture of NEMP than in a mixture of LW.   \\

\noindent
Suppose that $L_{\rm tube}=10\ \mu$m and $\delta c_{\rm a}=-c_{\rm ac}/10$.
If we adopt $G=3\ \mu$m$^4/$s from Fig.~\ref{fig:osmcoeff}(b),
\textcolor{black}{E}q.~(\ref{eqn:flrate}) \textcolor{black}{becomes equal to} $0.03\ \mu$m$^3/$s.
Because {of} $r_{\rm tube}=0.1\ \mu${m}, the average of $v_z$ across the tube's cross-section
is approximately $1\ \mu$m$/$s, which could be detected experimentally.
\textcolor{black}{I}t is shown at the end of Appendix \ref{sec:prof} that the average for large $\tau$ and small $|h|$
can be approximately calculated {using} the Gaussian model.
As $\tau$ increases well beyond the range of Fig.~\ref{fig:osmcoeff}, 
$\xi$ near the tube wall approaches \textcolor{black}{a molecular size} 
and {the RHS of Eq.~(\ref{eqn:fromdiffo})} 
should stop increasing in proportion to $\tau^\gamma$
because of the background contribution to the
osmotic susceptibility \cite{sengers,swinney}. 
\textcolor{black}{Then, our coarse-grained description would fail and}
diffiusioosmosis should be explained in terms of
the conventional mechanism rather than the PA \cite{anders, mood, diffphore}.
\section{Summary and further discussion\label{sec:dis}}
Assuming that  the free-energy functional is given by Eq.~(\ref{eqn:general}), 
that the tube {is not necessarily cylindrical but}
has the same cross section anywhere, and that the mixture at equilibrium has 
either critical or near-critical composition
in the central region of the reservoir, 
we derive some properties on the Onsager coefficients, appearing in Eq.~(\ref{eqn:newlinph}), in Section \ref{sec:sol}.
As is well expected, the pressure difference $\delta P$ causes the total mass flow rate
with a mixture flowing out of the tube being rich in the adsorbed component.  
The mass-fraction difference $\delta c_{\rm a}$
can cause the total mass flow, {\it i.e.\/}, diffusioosmosis emerges, only in the presence of PA.
The Onsager coefficients ${\cal L}_{21}$ and ${\cal L}_{12}$, which are equal to each other
owing to the reciprocal relation,  
are involved with these effects of PA, respectively. 
The coefficient ${\cal L}_{22}$ is separated into two parts;
${\cal L}_{22{\rm v}}$ involves the convection and ${\cal L}_{22{\rm d}}$ involves the interdiffusion. 
The former becomes nonzero only in the presence of PA.
\\

\noindent
In the presence of PA, the homogeneity at equilibrium is broken, and
the first two terms on the RHS of Eq.~(\ref{eqn:stok}) cannot be unified into a gradient of some scalar pressure.
This makes ${\cal L}_{12}={\cal L}_{21}$ and ${\cal L}_{22{\rm v}}$ nonzero. Thus, these 
effects of PA {comes} from the reversible part of the pressure tensor, $\Pi_{\rm rev}$.
The transport coefficients are introduced in relating irreversible fluxes linearly with thermodynamic forces defined in the bulk of a fluid;
$\eta_{\rm s}$ relates the viscous stress with the rate-of-strain tensor and $\Lambda$ relates
the difference between the diffusion fluxes of the components with the difference between the chemical potentials \cite{gromaz, miya}.
Whether the PA is present or not,  Curie's principle 
prohibits the viscous stress and the diffusive flux from being linearly related with the gradient of chemical potential
and with the velocity gradient, respectively \cite{gromaz}.  The prohibition can be also explained
by the difference in the parity concerning the time-reversal symmetry between ${\v v}$ and $\varphi$ \cite{gard}.
\\

\noindent
To evaluate the effects of PA quantitatively, we
use the free-energy functional coarse-grained in the renormalized local functional theory \cite{OkamotoOnuki}.
{W}e calculate the Onsager coefficients by assuming the tube to be cylindrical and assuming the mixture at equilibrium 
to have the critical composition in the central region of the reservoir;
their expressions are shown in Section \ref{sec:formulae}.
{In this model, $\delta P$ is not always equal to $\delta p$, as mentioned at Eq.~(\ref{eqn:dP}).}
In the absence of PA, the
Hagen-Poiseulle flow is generated 
{u}nder the pressure difference ($\delta P\ne 0$ and $\delta c_{\rm a}=0$), with
the interdiffusion being  usually accompanied, whereas
the interdiffusion without the total mass flow is generated by the mass-fraction difference 
($\delta P=0$ and $\delta c_{\rm a}\ne 0$). 
The effects of PA mentioned in the first paragraph of this section are numerically confirmed; 
the dimensionless Onsager coefficients,
introduced at Eq.~(\ref{eqn:LcalL}), 
are plotted in Figs.~\ref{fig:lutwat}(a), \ref{fig:lutwatvisc}(c), \ref{fig:viscoMPNE}(b), and \ref{fig:viscoMPNE}(c), whereas
the dimensionless mass flow rates, introduced at Eq.~(\ref{eqn:dimmass}), are plotted in Figs.~\ref{fig:lutwatmnr}
and \ref{fig:mpnemnr}.
In the diffusioosmosis due to PA,
{t}he total mass of a mixture flows into the reservoir {with a smaller composition of the preferred component}, as in the conventional diffusioosmosis \cite{anders,osmflow}.  {However},  
their mechanism{s are different},
as mentioned in the fourth paragraph of Section \ref{sec:intro} and at the beginning of Section \ref{sec:diffusio}. 
The conductance in diffusioosmosis due to PA, introduced at Eq.~(\ref{eqn:flrate}),
exhibits a non-monotonic change with a peak as plotted against $\tau$ (Figs.~\ref{fig:osmcoeff}(b) and (d)),
owing to the opposing effects of the osmotic susceptibility on the conductance mentioned below Eq.~(\ref{eqn:flrate}). 
{T}he critical enhancement of the transport coefficients {i}s evaded by the PA.  
\\

\noindent
In the calculation above, the difference in the parentheses of Eq.~(\ref{eqn:st3xy+}) equals $\psi^{(0)}$.
{For our results of the Onsager coefficients to be meaningful for $h\ne 0$,}
$\psi^{(0)}$ is required to be much larger than $\varepsilon \psi^{(1)}$ {somewhere} in the tube. 
Considering that $\psi$ is changed from $\psi^{(0)}$ by the velocity field,
this requirement can be simplified in terms of a typical energy per unit area as
$\eta_{\rm s} U \ll \left|h\psi^{(0)}(r_{\rm tube})\right|$, where
$U$ is a typical speed of the velocity field and can be taken as the average of $v_z$ across
the tube's cross-section.  The RHS of this inequality is larger than approximately 
$10^{-6}\ $kg/s$^2$ in Fig.~\ref{fig:lutwatvelpsi}(a), where
${\hat \psi}^{(0)}$ is calculated with some sets of values of $\tau$ and $h$ for a mixture of LW;  
this value of the lower bound is obtained from the value of ${\hat \psi}^{(0)}(1)$ 
for $\tau=4.0\times 10^{-4}$ and $h=0.01\  $cm$^3$/s$^2$.  
Thus, for the sets of values,
the requirement is met when the average of $v_z$ is much smaller than approximately $400\ \mu$m/s.
This value is approximately the same for a mixture of NEMP under the same sets of values of $\tau$ and $h$
although data not shown. 
Then, the Reynolds number, with $r_{\rm tube}=10^{-1}\ \mu$m 
taken as a typical length, is much smaller than unity,
which is also required in the linear regime.  
Considering ${\hat\psi}^{(0)}(1)$ can be roughly approximated  to be unity in Fig.~\ref{fig:lutwatvelpsi}(a)
we can nondimensionalize the typical speed by dividing it by $h\psi_*/\eta_*$ to use the quotient as the smallness parameter.
Judging from the velocity profile shown in Fig.~\ref{fig:lutwatvelpsi}(b), an estimate of $U$
in flow driven by the pressure difference is given by the average of $v_z$ in the Hagen-Poiseulle flow.
An estimate of $U$ in flow driven by the mass-fraction difference  
is given by Eq.~(\ref{eqn:slipvel}) when the approximation $\omega\approx \tau$ is valid and 
$\xi$ at the critical composition is much smaller than $r_{\rm tube}$.
\\

\noindent
Changes of the Onsager coefficients are continuous
with respect to $\tau$ in our numerical \textcolor{black}{results.}
{When the mass fraction of the component disfavored by the tube wall is 
larger than its critical value at the central region of the reservoir}
in the reference state,
the capillary condensation transition can occur in the tube at a value of $\tau$
 \cite{OkamotoOnuki,capcondens}. 
If this transition occurs as a first-order transition, the change in the Onsager coefficients should be discontinuous
with respect to $\tau$ and exhibit hysteresis.  Similar transitions have been reported for ion-containing mixtures 
\cite{tsori2007,tsori2011,tsori2016, precip, precip2}.   It remains to be studied how the transport properties of a mixture in a tube 
change when it undergoes this kind of transition.
The mass flow rates caused by
the temperature difference between the reservoirs are also of interest. 
The renormalized local functional theory should be extended for this study 
 so that it can describe 
the equilibrium density of a mixture's   {entropy}. \\
{
\section{Conclusions}
In} the isothermal transport of a mixture through a tube with the PA,
  the convective flow rich in the preferred component
caused by {the} {pressure difference is related via cross-effect Onsager coefficients
with the total mass flow caused by} {the} {mass-fraction difference.
The resulting diffusioosmosis has a mechanism involving the near-criticality,
like diffusiophoresis in a near-critical mixture discussed recently by one of the present authors \cite{diffphore}. 
According to our numerical results, the critical enhancement of the transport coefficients is evaded by the PA,
the conductance in diffusioosmosis 
changes non-monotonically with respect to the temperature, and the change is
large enough to be detected experimentally.   }

\section*{Acknowledgments}
The authors thank Prof.~P. G. Wolynes for informing them of Ref.~\cite{wolynes}. 
YF appreciates stimulating 
discussion with T. Iyori. S. Y. was supported by JSPS KAKENHI Grant-in-Aid for Young Scientists (B) (15K17737 and 18K13516).

\appendix
\begin{section}{Non-dissipative part of the stress tensor\label{sec:stress}}
As in conventional thermodynamics, we can derive the reversible part of the pressure tensor
by considering the change of ${F}$ due to an infinitesimal quasistatic deformation of a mixture \cite{okafujiko, onukibook}.
In this non-dissipative deformation, the chemical potentials $\mu_{\rm a}$ and $\mu_{\rm b}$
can be inhomogeneous, like an inhomogeneous magnetic field in a magnetic system.
This means that each locus in a mixture has each particle bath.  
Here, unlike in the text, $V_{\rm tot}$ is the deformable region occupied by the mixture, 
$t$ is not the time but a parameter of deformation,  ${\v v}$ is the
displacement of fluid particle per unit $t$, and  
${\v j}_n$ represents the flux to the particle bath. 
The last implies that ${\v j}_{\rm a}+{\v j}_{\rm b}$ does not always vanish.
Consider a small region, $V_t$, {co-moving} with the deformation.
This region is assumed to have the smallest volume that allows us to define $\mu_n$ and $\Pi_{\rm rev}$.
They can be regarded as homogeneous over {$V_t$}.
Below, the repeated indices are summed up.
The local equilibrium in the bulk gives
\begin{equation}\frac{d}{dt}\int_{V_t} d{\v r}\ f_{\rm bulk}=
\mu_n\ \frac{d}{dt}\int_{V_t} d{\v r}\ \rho_n({\v r}, t)-\Pi_{\rm rev} : \int_{\partial V_t} dA\ {\v n}_{\partial V_t}{\v v}
\ ,\label{eqn:loc}\end{equation}
where the symbol $:$ represents the double dot product. 
Equation (\ref{eqn:loc}) yields
\begin{equation}
\frac{Df_{\rm bulk}}{Dt}+f_{\rm bulk} \nabla\cdot {\v v}= -\mu_n \nabla\cdot{\v j}_n-\Pi_{\rm rev} : \left(\nabla{\v v}\right) 
\label{eqn:loceq}\end{equation} 
with the aid of Eq.~(\ref{eqn:alpdiffusion}).  Here, $D/Dt$ represents the Lagrangian ``time-derivative''. 
Because $V_{\rm tot}$ also co-moves with the deformation, 
we have
\begin{equation}
\frac{d{F}}{dt}
=-\int_{V_{\rm tot}}d{\v r}\ \left[ \mu_n \nabla\cdot{\v j}_n+\Pi_{\rm rev} : \left(\nabla{\v v}\right) \right]
\label{eqn:Fhenka}\ ,\end{equation} 
apart from a surface term, {which involves $f_{\rm surf}$}.  Here, $\mu_n$ and $\Pi_{\rm rev}$
can depend on ${\v r}$. \\

\noindent
We below outline how Eqs.~(\ref{eqn:murhophi}) and
(\ref{eqn:genPi}) are derived from Eq.~(\ref{eqn:general}).
The Lagrangian ``time-derivative'' of $f_{\rm bulk}$ is found to be the sum of 
\begin{equation}
\frac{\partial f_{\rm bulk}}{\partial \rho} \frac{D\rho}{Dt}+
\frac{\partial f_{\rm bulk}}{\partial \left(\nabla\rho\right)}\cdot \nabla\left(\frac{D\rho}{Dt}\right)
-\frac{\partial f_{\rm bulk}}{\partial \left(\nabla\rho\right)}\cdot \left(\nabla{\v v}\right)\cdot\left(\nabla\rho\right) 
\end{equation}
and the above with $\rho$ replaced by $\varphi$.
Substituting this result into the LHS  of Eq.~(\ref{eqn:loceq}), and integrating the result over $V_{\rm tot}$, we
can transform the contribution from the bulk to $d{F}/(dt)$ into the form of the RHS of Eq.~(\ref{eqn:Fhenka})
with $\mu_n$ and $\Pi_{\rm rev}$ given by Eqs.~(\ref{eqn:murhophi}) and
(\ref{eqn:genPi}), except for a new surface term, which is generated by the application of the divergence theore{m}.

\noindent
{B}elow,
$\nabla_\parallel \cdot {\v v}_\parallel$ indicates the divergence defined on 
$\partial V_{\rm tot}$ and
$H_{\rm m}$ denotes the mean curvature of $\partial V_{\rm tot}$.  The latter is defined so that
it is positive when the center of curvature lies on the side directed by ${\v n}_{\partial V_{\rm tot}}$.
The surface term unwritten in Eq.~(\ref{eqn:Fhenka}) is combined with the new surface term.
{T}he combination {gives} the sum of
\begin{equation}
\int_{\partial V_{\rm tot}} dA\ \frac{D\rho}{Dt} \left( \frac{\partial f_{\rm bulk}}{\partial \left(\nabla\rho\right)}\cdot
 {\v n}_{\partial V_{\rm tot}}+\frac{\partial f_{\rm surf}}{\partial \rho} \right),
\end{equation} the above with $\rho$ replaced by $\varphi$, and
\begin{equation}
\int_{\partial V_{\rm tot}} dA\ 
f_{\rm surf}\left( \nabla_\parallel \cdot {\v v}_\parallel -2H_{\rm m} {\v v}\cdot{\v n}_{\partial V_{\rm tot}} \right)\ .\label{eqn:surfaceint}
\end{equation}
We can use Eq.~(\ref{eqn:alpdiffusion}) to
rewrite the Lagrangian ``time-derivatives'' of the mass densitie\textcolor{black}{s}
in terms of ${\v j}_n$.  This flux
should not exist at $\partial V_{\rm tot}$, where the particle bath does not exist.  
This means that the coefficients of the ``time-derivatives'' should vanish there, which
leads to Eq.~(\ref{eqn:surfbound}).  {Thus,} this equation {holds} even in the dynamics \cite{DJ}.
Hence, the surface term unwritten on the RHS of Eq.~(\ref{eqn:Fhenka}) is reduced to Eq.~(\ref{eqn:surfaceint}),
which means that $-f_{\rm surf}$ works as a two-dimensional pressure in $\partial V_{\rm tot}$.
The force density $2f_{\rm surf}H_{\rm m} {\v n}_{\partial V_{\rm tot}}$ 
yields the Laplace pressure, as discussed in Appendix A of Ref.~\cite{effvis}. \\

\noindent
The reversible parts of the hydrodynamic equations contain nonlinear terms, 
as shown in Eqs.~(\ref{eqn:genPi}) and (\ref{eqn:diffusion}). 
The sum of the first two terms on the RHS of Eq.~(\ref{eqn:stok})
equals the sum of $-\rho_n\nabla\mu_n$ over $n=$a and b.
Each term  is linked with the first term on the RHS of Eq.~(\ref{eqn:alpdiffusion}) 
via the reciprocal relation after the hydrodynamic equations are linearized, as shown in Ref.~\cite{miya}.
This is expected to underlie the cross effects mentioned in \textcolor{black}{Section} \ref{sec:sol}.
\end{section}

\begin{section}{Reciprocal relation \label{sec:insep}}
We consider two sets of flow fields, each being driven by
the thermodynamic forces $\left(-\delta P_k/{T}, -\delta\mu_k/{T} \right)$, 
with $k$ being \textcolor{black}{i or ii. 
The resultant} thermodynamic fluxes
and fields in the tube are also indicated by the subscript $_k$. 
Different ways of applying the divergence theorem
to the volume integral of $\eta_0 E^{(1)}_{{\rm i}} : E^{(1)}_{\rm ii}$ over the tube {interior}, denoted by $V_{\rm tube}$,
yield
\begin{equation}
\int_{V_{\rm tube}}d{\v r}\ {\v v}_{\rm ii}^{(1)} \cdot \left[\nabla\cdot\left({\eta}_0 E_{\rm i}^{(1)}\right)\right]
=\int_{V_{\rm tube}}d{\v r}\ {\v v}_{\rm i}^{(1)} \cdot \left[\nabla\cdot\left({\eta}_0 E_{\rm ii}^{(1)}\right)\right]
\ ,\label{eqn:EE}\end{equation}
where we note that the no-slip boundary condition is imposed at the \textcolor{black}{tube} \textcolor{black}{wall}, 
\textcolor{black}{that} $\nabla\cdot {\v v}^{(1)}_k$ vanishes in a stationary laminar flow in the \textcolor{black}{tube, and that 
effects of the tube edges are assumed to be negligible.}
Substituting Eq.~(\ref{eqn:stok}) into Eq.~(\ref{eqn:EE}), we use Eq.~(\ref{eqn:nablamuP}) to find that  
\begin{equation}
\frac{\delta P_{\rm i}-\varphi_{\rm c}\delta\mu_{\rm i}}{\rho^{({\rm ref})}L_{\rm tube}}  \int_{S_{\rm tube}}dA\ \rho^{(0)} v_{{\rm ii}z}^{(1)}
+\frac{\delta\mu_{\rm i}}{L_{\rm tube}} \int_{S_{\rm tube}}dA\ \varphi^{(0)}  v_{{\rm ii}z}^{(1)}
\end{equation}
equals the above equation with the subscripts $_{\rm i}$ and $_{\rm ii}$ exchanged.
Putting $\delta \mu_{\rm i}$ and $\delta P_{\rm ii}$ equal to zero, we have
{$
  {\cal I}_{\rm ii} \delta P_{\rm i}=
{\cal J}_{\rm i} \delta\mu_{\rm ii}$} because $j_{{\rm i}z}$ vanishes.
{Thus,} ${\cal L}_{12}={\cal L}_{21}$ {is derived.} 
{This} derivation {remains} valid even when $\rho^{(0)}$ is inhomogeneous, unlike
{that of} Ref.~\cite{xu}.
\end{section}

\begin{section}{Renormalized local functional theory\label{sec:rlft}}
The composition profile {in a mixture} has the probability density
in equilibrium fluctuations.
The effective Hamiltonian is defined in such a way that
the exponential of its negative is proportional to the probability density functional. 
{As a bare model for the bulk of a mixture, we can} {assume}  {the effective Hamiltonian of}
the Ginzburg-Landau-Wilson type, or the $\psi^4$ model, {for the order-parameter}.
{N}ear the critical point, the bare model gives nearly the same probabilities, close to the maximum probability,
 to many composition profiles, which differ from each other only by details with length scales smaller than the local
correlation length, and {hence} 
remarkable composition fluctuations arise over these scales.  Changing the way of counting profiles to neglect
fine differences between profiles,
we can unify many profiles into much fewer profiles.  That is, coarse-graining $\psi$ up to
the local correlation length enables us to {regard} the average profile as 
maximizing the resultant probability density functional, which is called renormalized local functional \cite{OkamotoOnuki}. 
The $\psi$-dependent part of the coarse-grained effective Hamiltonian for 
the bulk of a mixture with $\mu=0$ is given by the volume integral of the sum of 
\begin{equation}
\frac{1}{2} C_1\xi_0^{-2}\omega^{\gamma-1}\tau\psi^2 +
\frac{1}{12} C_1C_2\xi_0^{-2} \omega^{\gamma-2\beta} \psi^4 \ ,\label{eqn:rlftinteg}
\end{equation}  and 
$C_{1}\omega^{-\eta\nu} \left\vert\nabla\psi\right\vert^2/2$
{o}ver $V_{\rm tot}$.
Here, $C_1$ is a positive 
nonuniversal constant {and} $C_{2}$ is given by
$C_{2}=3u^{*}C_{1}\xi_{0}$.
The quantity $\omega$, defined by Eq.~(\ref{eqn:xiomega}), satisfies
a self-consistent condition, 
$\omega=\tau+C_{2}\omega^{1-2\beta}\psi^{2}$, 
{b}ecause the mean-field approximation is valid after coarse-graining up to $\xi$. 
The logarithm of the probability density  is proportional to Eq.~(\ref{eqn:relation})
divided by $-k_{\rm B}T$.  Thus, we can identify {$f_-(\psi)-(\mu_-)_{\rm c}\varphi$} 
with the product of $k_{\rm B}T$ and
Eq.~(\ref{eqn:rlftinteg}), and
identify $M_-(\psi)$ with $k_{\rm B}TC_1\omega^{-\eta\nu}$.\\

\begin{figure}
\includegraphics[width=6cm]{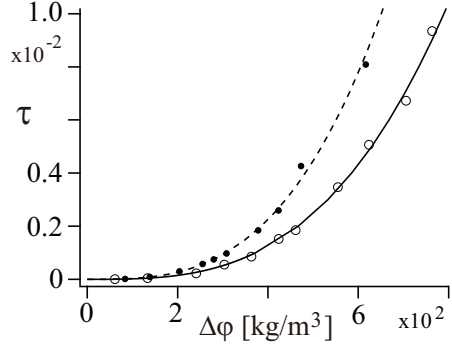}
\caption{
Relationship between $\tau$ and the difference in $\varphi$ between coexistence phases, denoted by
$\Delta\varphi$, in a mixture.
The open and solid circles are replots of data in Ref.~\cite{to} for a mixture of LW
and data in Ref.~\cite{wims} for a mixture of  NEMP, respectively.
The solid and dashed curves are obtained by the curve-fits of twice the value of
Eq.~(3.3) of Ref.~\cite{OkamotoOnuki} to these sets of data, respectively. }
\label{fig:jayacoex}
\end{figure}
\noindent
{Estimates of $C_2$ are obtained as follows.} 
The linear relationship between the refractive index 
and the volume fraction in a mixture of LW {are} shown in Fig.~2 of Ref.~\cite{to}.
Using the three values of the circles in this figure and
the index of the pure water, $1.330$, we obtain the fitted line with  
the slope of $1.77\times 10^{-1}$.
Reference \cite{to} also shows the refractive indices of the coexistent two phases above the lower consolute temperature.
Converting the difference \textcolor{black}{between} the indices into \textcolor{black}{the one in} $\varphi$ with the aid of the 
values of ${\bar v}_n^{({\rm ref})}$ mentioned in the text, we plot the relationship between
the latter difference $(\Delta\varphi$) and $\tau$ in Fig.~\ref{fig:jayacoex}.
The coexistence curve for $\mu=0$ is given by Eqs.~(3.3) and (3.23) of Ref.~\cite{OkamotoOnuki}, and the data
are approximately regarded as obtained for $\mu=0$. 
{The} curve-fit to the {data} 
yields $C_2=7.14\times 10^{-7}\ $m$^6$/kg$^2$ for a mixture of LW.
{For} a mixture of NEMP, we obtain the relationship between
$\Delta\varphi$ and $\tau$ from data (Run A) of Ref.~\cite{wims} and find $C_2=1.05\times 10^{-6}\ $m$^6$/kg$^2$ 
(Fig.~\ref{fig:jayacoex}).
\end{section}

\begin{section}{{Composition profiles and flow profiles \label{sec:prof}}}
{Applying the procedure mentioned in the last paragraph of Section \ref{sec:hyd}, we find that}
${\hat \psi}^{(0)}({\hat r})$ is the solution of
the differential equation for ${\hat\psi}({\hat r})$,
\begin{equation}
0={\hat f}'({\hat\psi})-\frac{1}{2}\frac{d {\hat \omega}^{-\eta\nu}}{d{\hat\psi}}
\left(\partial_{\hat r}{\hat\psi}\right)^2-{\hat \omega}^{-\eta\nu}\left(\partial_{\hat r}^2+
\frac{1}{{\hat r}}\partial_{\hat r}\right){\hat\psi}\quad {\rm for}\ {\hat r}<1\ ,\label{eqn:stationary}
\end{equation} {where 
 ${\hat \omega}$
is regarded as a function of ${\hat \psi}$ via Eq.~(\ref{eqn:omega2})}, 
together with the boundary condition,
$\partial_{\hat r}{\hat\psi}={\hat h} {\hat \omega}^{\eta\nu}$ at ${\hat r}=1$, {with}
${\hat h}$ {being defined} as 
${h T^* /\left(T\mu_*r_{\rm tube}\right)}$.
These equations indicate that {the sign of ${\hat \psi}^{(0)}$
is changed by changing that of $h$.}
We use
{these equations and Eq.}~(\ref{eqn:vzprofile2}) to obtain
Fig.~\ref{fig:lutwatvelpsi} for a mixture of LW.  
Because $h>0$ is assumed,
${\hat \psi}^{(0)}$ is positive and increases toward the wall in Fig.~\ref{fig:lutwatvelpsi}(a).  
As $\tau$ decreases for $h=0.1\ $cm$^3/$s$^2$, ${\hat\psi}^{(0)}$
increases throughout the tube, which can be interpreted as thickening of
the adsorption layer.  
{U}sing ${\hat \psi}^{(0)}({\hat r})$ for ${\hat \psi}$ in Eq.~(\ref{eqn:omega2}),
we calculate ${\hat \xi}$, defined as $\xi/r_{\rm tube}$, with the aid of Eq.~(\ref{eqn:xiomega}).
Its smallest value in the inset of Fig.~\ref{fig:lutwatvelpsi}(a) {gives $\xi=3\ $nm,
which} is sufficiently large as compared with a molecular size.  
{This} is required in our coarse-grained description.
\\

\noindent
{B}y setting $(\delta{\hat P},\delta{\hat\mu})=(-\varepsilon,0)$ {for a mixture of LW},  
we calculate $v_z^{(1)}$ and plot the results in  Fig.~\ref{fig:lutwatvelpsi}(b), {where ${\hat v}_z^{(1)}$ is a dimensionless quantity
defined as $v_z^{(1)}$} divided by $8{\cal I}_* /({\pi r_{\rm tube}^2}){= 2.64}\ \mu$m/s. 
The {quantity} decreases with $\tau$, which is consistent with the increase of $\eta_{\rm s}$ in Fig.~\ref{fig:lutwatvisc}(a).
Owing to the slight change of $\eta_{\rm s}$, {the quantity} 
does not so much deviate from the {black} dotted curve, which
is obtained by setting ${\hat\eta}$ to unity and represents the Hagen-Poiseuille flow, {${\hat v}_z^{(1)}=(1-{\hat r}^2)/4$}.
Setting $\delta{\hat P}$ as above
 and $\delta{\hat \mu}$ equal to not zero but $-{\hat {\cal L}}_{11}\delta {\hat P}/{\hat {\cal L}}_{12}=\varepsilon 
{\hat {\cal L}}_{11}/{\hat {\cal L}}_{12}$ to make ${\hat {\cal I}}$ vanish at $\tau=1.25\times 10^{-5}$, 
we calculate ${\hat v}_z^{(1)}$ and plot the result with the  {red} thicker dash-dot-dot curve in Fig.~\ref{fig:lutwatvelpsi}(b).
{It is positive near the tube center}, {and negative in an outer region}, {   
where the effect} of $\delta {\hat \mu}$ exceeds
{that of} $\delta {\hat P}$.  
{T}his kind of bidirectional flow is also pointed out in Ref.~\cite{roij}.
In Fig.~\ref{fig:lutwatvelpsi}(c),
${\hat v}_z^{(1)}$ for $(\delta{\hat P},\delta{\hat\mu})=(0, -\varepsilon)$
increases as the adsorption is stronger. 
{A}t the  {largest} value of $\tau$, 
the cylindrical shape of the tube has little effect on the convection,  
and ${\hat v}_z^{(1)}$ appears to change only near the tube wall, or in other words,
appears to slip between the mixture bulk and the wall.
\\

\begin{figure}
\includegraphics[width=12cm]{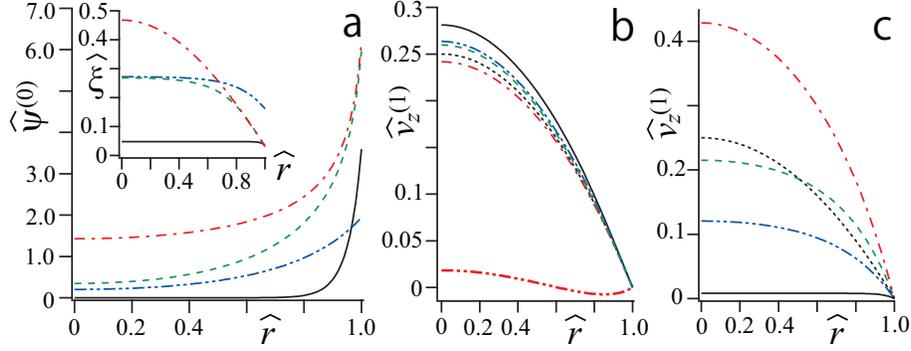}
\caption{Calculation results for a mixture of LW.
Against ${\hat r}$, the dimensionless order parameter ${\hat\psi}^{(0)}({\hat r})$ is plotted in (a),
the dimensionless correlation length
${\hat \xi}$ at equilibrium is plotted in the inset of (a), 
${\hat v}_z^{(1)}$ for $(\delta{\hat P},\delta{\hat\mu})=(-\varepsilon,0)$
is plotted in (b), and ${\hat v}_z^{(1)}$ for $(\delta{\hat P},\delta{\hat\mu})=(0,-\varepsilon)$
is plotted in (c), {except for} 
the red thicker dash-dot-dot curve  {at the bottom} in (b) and the black dotted curves in (b) and (c).
{See the second paragraph of Appendix \ref{sec:prof} for these exceptions and
the definition of ${\hat v}_z^{(1)}$. } 
The red dash-dot, green dashed, and black solid curves are obtained 
for $\tau=1.25\times 10^{-5}$, $4.0\times 10^{-4}$, and $6.4\times 10^{-3}$, respectively,
under $h=0.1\ $cm$^3/$s$^2$.
The blue dash-dot-dot curves are obtained for $\tau=4.0\times 10^{-4}$ and  $h=0.01\ $cm$^3/$s$^2$.
}
\label{fig:lutwatvelpsi}
\end{figure}

\noindent
Considering that ${\hat \xi}>0.44$ holds for ${\hat r}<0.22$ 
at $\tau=1.25\times 10^{-5}$ in the inset of Fig.~\ref{fig:lutwatvelpsi}(a),  
the spatial resolution of our coarse-grained description in this region near the tube axis
is approximately \textcolor{black}{$0.44r_{\rm tube}$}.
Each of the variations shown by the curves for ${\hat r}<0.22$ at $\tau=1.25\times 10^{-5}$ 
in Figs.~\ref{fig:lutwatvelpsi}(b) and (c) is much smaller than each of the value at ${\hat r}=0$,
indicating that we can trace each velocity profile well on the   
spatial resolution mentioned above.  \textcolor{black}{Each velocity profile exhibits a more rapid change 
near the tube wall, which can be traced on the local resolution estimated from ${\hat \xi}$,
which is smaller there than in \textcolor{black}{the region near the tube axis}.
Similar discussions are valid for the other curves in Fig.~\ref{fig:lutwatvelpsi}. 
Thus, our coarse-grained description is valid for the velocity fields examined in this figure. }\\

\noindent
Suppose that the second term is much smaller than the first term in Eq.~(\ref{eqn:omega2}).  Then,
${\hat \omega}\approx {\hat \tau}$ holds and the second term in the parentheses
of Eq.~(\ref{eqn:prefmm}) is negligible.  We \textcolor{black}{further} approximate $M_-$
to be $k_{\rm B}T_{\rm c} C_1$ to obtain   
the Gaussian model, which is valid for large $\tau$ and small $|h|$. 
We here consider a simple case, where the Gaussian model is valid, $\eta_{\rm s}$ is constant, and \textcolor{black}{$r_{\rm tube}$} is so large
that the tube wall can be regarded as a plane. 
Far from the wall \textcolor{black}{in flow driven by the mass-fraction difference, $v_z$ is given by}
\begin{equation}
\frac{\delta\mu}{\textcolor{black}{\eta_{\rm s}} L_{\rm tube}}  \int_0^\infty dX_1\ \int_\infty^{X_1} dX\ \psi^{(0)}(X) 
\label{eqn:slip}\end{equation}
\textcolor{black}{owing to} Eq.~(\ref{eqn:st3xy+}).  Here, 
$\psi^{(0)}$ is regarded as a function of the distance from the tube wall, $X$.
From the equations corresponding with {Eq}.~(\ref{eqn:stationary}) and {the associated boundary condition}
we have $\psi^{(0)}(X)=h\xi e^{-X/\xi}/M_-=\xi^2 {\psi^{(0)}}''(X)$, where $\xi$ is homogeneous \cite{EPJ}.  Thus, the apparent slip velocity,
\textcolor{black}{given by} Eq.~(\ref{eqn:slip}), \textcolor{black}{becomes}
\begin{equation}
-\xi^2\psi^{(0)}(0) \frac{\delta\mu}{\eta_{\rm s}L_{\rm tube}} \approx -2h \xi \rho_{\rm c}\frac{\delta c_{\rm a}}
{\eta_{\rm s}L_{\rm tube}}
\label{eqn:slipvel}\end{equation}
because of ${\bar v}_-^{({\rm ref})}\varphi_{\rm c}\ll 1$.
Using $\xi_0\tau^{-\nu}$ for
the homogeneous value of $\xi$ and
substituting the parameter values of the black solid curve in Fig.~\ref{fig:lutwatvelpsi}(c)
into the LHS above, 
we find the value of ${\hat v}_z$ corresponding with the apparent slip velocity to be $0.008$,
which agrees with the value shown by the flat part of the black solid curve. 
For $\delta c_{\rm a}=-c_{\rm ac}/10$, $L_{\rm tube}=10\ \mu$m,  $\tau=6.4\times 10^{-3}$ and $h=0.01\ $cm$^3/$s$^2$,
the RHS of Eq.~(\ref{eqn:slipvel}) is approximately $0.1$ and $0.9\ \mu$m$/$s
for a mixture of LW and a mixture of NEMP, respectively.   
They agree with the averages of $v_z$
over tube's cross-section calculated from the values of asterisks on the extreme \textcolor{black}{right} in Figs.~\ref{fig:osmcoeff}(b) and (d),
respectively.  {In passing, the double integral of Eq. (D2) becomes proportional to $\tau^{\nu-\beta-\gamma}$ if we use the universal profile in the critical regime \cite{fisher-degennes,diehl97}.}
\end{section}

\begin{section}{Transport coefficients\label{sec:traco}}
We first mention the previous studies of
the dynamic renormalization group calculation 
for the bulk of a mixture \cite{sighalhoh, onukishear, folkmos}. 
The $\psi^4$ model is used as a bare model for
an incompressible mixture having vanishing average of $\psi$. 
In this mixture, $\xi$
is homogeneously equal to $\xi_0\tau^{-\nu}$.
The interdiffusion coefficient equals $\Lambda$ divided by the osmotic susceptibility,
given by $1/f_-''(0)$, and is found to
equal $\lambda^*k_{\rm o}^{4-z_\psi}{\xi}^{2-z_\psi}$ in a near-critical mixture. 
Here, $\lambda^*$ is a 
constant determined at the fixed point of the renormalization group flow starting from
a bare model having $k_{\rm o}$ as the cutoff wavenumber; $1/k_{\rm o}$ is a microscopic length scale.
With $z_\eta$ denoting $z_\psi-d$, where $d$ is the spatial dimensions,  
the singular part of the viscosity, denoted by $\eta_{\rm sing}$,
becomes given by $\eta^*(k_{\rm o}\xi)^{z_\eta}$ as the critical point is approached.
Here, $\eta^*$ is a constant determined at the fixed point and is not $\eta_*$ defined in the text.
{Thus, in equilibrium order parameter fluctuations at the critical composition, $\Lambda$ equals}
\begin{equation}
\frac{\lambda^*(k_{\rm o}{\xi})^{4-z_\psi}}{f_{-}''(0)\xi^2}=\frac{Rk_{\rm B}{T}_{\rm c}}
{f_{-}''(0)\xi^{d-2}\eta_{\rm sing}}  
\ ,\label{eqn:equL}\end{equation}
where $f_{-}''(0)$ is given by Eq.~(\ref{eqn:fprpr})
and $R$ is a universal constant called Kawasaki amplitude.  The {e}quality above comes from the definition of this constant.
Below, $d$ is put equal to three, as in the text. 
From the experimental data of $2\Lambda M_-/{\xi}^4$ \textcolor{black}{at the critical composition} \cite{mirz,iwan},
the value of $2\lambda^*k_{\rm o}^{4-z_\psi}\xi_0^{-z_\psi}$ is $2.5\times 10^{10}\ $s$^{-1}$
for a mixture of LW, and $1.23\times 10^{11}\ $s$^{-1}$
for a mixture of NEMP.  Applying these values for the first equality of Eq.~(\ref{eqn:equL}),
we obtain $\Lambda_*=2.19\times 10^{-9}\ $kg$\cdot$s$/$m$^3$
and $2.78\times 10^{-8}\ $kg$\cdot$s$/$m$^3$ for respective mixtures.
\\

\noindent 
The product $\nu z_\eta$
is measured to be around $0.042$ \cite{bergmold, bergmold2}, which leads to $z_\psi=3.067$.
Because of the weak singularity, the viscosity at the critical composition
can be well described in the form of multiplicative anomaly as \cite{ohta3,sengers}
\begin{equation}
A_{\eta}\exp{\left[\frac{B_{\eta}}{T-T_{\eta}}\right]} \exp{\left[z_\eta H(\tau, q_{\rm D},q_{\rm C})\right]}\ ,
\label{eqn:multi}\end{equation}
where $H(\tau, q_{\rm D},q_{\rm C})$ is the crossover function with $q_{\rm D}$ and $q_{\rm C}$ being wavenumber parameters. 
It is defined at Eq.~(2.18) of Ref.~\cite{bhatt}, approaches zero as $\xi$ tends to zero, and approaches the sum of $\ln{\xi}$ and a constant 
as $q_{\rm C}{\xi}$ is larger.  Thus, $\eta_{\rm sing}$
is given by Eq.~(\ref{eqn:multi}) with $T-T_\eta$ being replaced by $T_{\rm c}-T_\eta$.
This leads to $\eta_{\rm sing}\propto  \xi^{z_\eta}$ for $T\approx T_{\rm c}$, as it should do.
We approximate $\eta_*$ to be Eq.~(\ref{eqn:multi}) at $T=T_*$, 
{c}onsidering that $\tau_*$ is very small.
The calculation result for $\Lambda$ in the mode coupling theory \cite{kawasaki}, 
agrees with the RHS of Eq.~(\ref{eqn:equL}) if
$R$ is put equal to $1/(6\pi)$ and if $z_\eta$ is regarded as zero.
\\

\noindent
In Eqs.~(\ref{eqn:diffusion}) and (\ref{eqn:stok}), we 
need the values of the transport coefficients for off-critical compositions.
As in Eq.~(4.11) of Ref.~\cite{yabufuji}, we obtain the value of $\Lambda_0$ by regarding $\xi$ as given 
from ${\hat \psi}^{(0)}$ via Eqs.~(\ref{eqn:xiomega}) and {(\ref{eqn:omega2})}
and replacing $f_{-}''(0)$ with  $f_{-}''(\psi^{(0)})$ in 
{the LHS of} Eq.~(\ref{eqn:equL}).  
This definition yields Eq.~{(\ref{eqn:hatLamdef})}. 
{A similar function for $\Lambda_0$ is used in Sec.~4.2 of Ref.~\cite{yabufuji}, 
where $z_\psi$ is put equal to three.}
The fraction in the last term in the last entry of Eq.~(\ref{eqn:calLcomp1}) is originally
the LHS of Eq.~(\ref{eqn:fraction}) below.
We use the definitions to obtain 
\begin{equation}
\frac{2\pi r_{\rm tube}^2}{L_{\rm tube}} \times\frac{\Lambda_*{T_*}\mu_*}{{\cal I}_*{T}\psi_*}
=\frac{48 T_{\rm c} u^* R}{{T}}
\ .\label{eqn:fraction}\end{equation}
Approximating {$T$} to be $T_{\rm c}$ and using $R=1/(6\pi)$, we obtain
the fraction in Eq.~(\ref{eqn:calLcomp1}).  \\

\noindent
We below mention \textcolor{black}{$\eta_{\rm s}$} for off-critical compositions.
As a fitting function to the data out of the peak in Figs.~\ref{fig:viscosity}(a) and \ref{fig:viscoMPNE}(a), we use
\begin{equation}
A_{\eta}\exp{\left[\frac{B_{\eta}}{T-T_{\eta}}\right]}+ G(c_{\rm a}-c_{\rm ac}) \ ,\label{eqn:regular}\end{equation}
where $G$ is a quartic function vanishing at $c_{\rm a}=c_{\rm ac}$ \cite{bhatt}.  The difference in the parentheses
in the \textcolor{black}{second} term above represents the variable of $G$.
The dependence of the four coefficients of the quartic function on $\tau$ is estimated from curve {fitting}.
Subtracting the background viscosity given by Eq.~(\ref{eqn:regular}) from the data, we fit the difference to \cite{tsai}
\begin{equation}
A_{\eta}\exp{\left[\frac{B_{\eta}}{T-T_{\eta}}\right]}\left\{
\exp{\left[z_\eta H(\tau, q_{\rm D},q_{\rm C})\right]}-1\right\}
\exp{\left[-\zeta \left(c_{\rm a}-c_{\rm ac}\right)^2\right]}
\ ,\label{eqn:singular}\end{equation}
where $\zeta$ is a positive constant to be estimated from the data in the peak.
Thus, we can obtain the viscosity as a function of $\tau$ and $c_{\rm a}$, which 
can be converted {to} $\varphi$ using
{Eq.~(\ref{eqn:rhovv})}. 
The function $\eta_0$ is obtained by replacing $\varphi$ with $\psi^{(0)}(r)+\varphi_{\rm c}$.
\\

\noindent
For a mixture of LW at the critical composition
immediately below the lower consolute temperature, $A_{\eta}$, $B_\eta$, $T_\eta$, $q_{\rm C}$, and $q_{\rm D}$ are estimated 
to be $1.4\times 10^{-7}\ {\rm Pa}\cdot{\rm s}$, $2.916\times 10^3\, {\rm K}$, $0.2\, {\rm K}$,
$10^{11} \,{\rm m}^{-1}$, and $10^{9}\ {\rm m}^{-1}$, respectively \cite{mirz}.  
Using the experimental data for off-critical compositions in Ref{s}.~\cite{gratt,stein}, we find 
$5.37+ 58.9\tau$,  $1.34+ 68.3\tau$, 
$-13.1 + 71.4\tau $, and  $-3.29-683\tau\ {\rm mPa}\cdot s$
as the coefficients of the linear, quadratic, cubic, and quartic term{s} of $G$,  
respectively.  From  the data shown by the circles in the peak in Fig.~\ref{fig:viscosity}{(a)}, we
obtain $\zeta=2.60\times 10^{2}$.   
{E}quation (\ref{eqn:multi}) with $q_{\rm C}=1.3\times 10^9\ $m$^{-1}$
and $q_{\rm D}=3.2\times 10^8\ $m$^{-1}$ 
is used for a mixture of NEMP \textcolor{black}{at the critical composition}
immediately above the upper consolute temperature in Ref.~\cite{iwan}.
We use the dashed line in Fig.~1 of this reference as 
the first term on the RHS of Eq.~(\ref{eqn:regular}).  
The experimental data in Ref.~\cite{leis} include
the values of $\eta_{\rm s}$ for off-critical compositions.  However, the data at the critical composition
are slightly different from  
those calculated from Eq.~(\ref{eqn:multi}) with the values of $q_{\rm C}$ and $q_{\rm D}$ mentioned above.  
Thus, so that they agree with each other, 
we shift the data  for  each value of $\tau$ in Ref.~\cite{leis}.
The sets of data at $\tau=1.47\times 10^{-4}$, $1.82\times 10^{-3}$, and $1.19\times 10^{-2}$
are respectively raised by $8.60\times 10^{-3}$, $4.80\times 10^{-3}$, and
$1.25\times 10^{-2}\ {\rm mPa}\cdot {\rm s}$.   The {results} are shown  in Fig.~\ref{fig:viscoMPNE}(a).
Then, we obtain 
$0.303-2.74 \tau$,  $-0.812+ 32.8\tau$, 
$0.212 + 13.5\tau $, and  $5.28-159\tau\ {\rm mPa}\cdot {\rm s}$
as the coefficients of the linear, quadratic, cubic, and quartic term{s} of $G$,  
respectively{,} and obtain $\zeta=2.66\times {10^{2}}$.  
\\

\noindent
{T}he life time of a correlated cluster with the size of $\xi$
in the order-parameter fluctuations would be given by the inverse of $\Lambda f''_-(0)/\xi^2$.
If the inverse of the shear rate of a flow is smaller than the life time, the cluster is deformed and the 
{c}ritical enhancement 
of the transport coefficients $\Lambda$ and $\eta_{\rm s}$ 
is suppressed \cite{onukishear, onukirev, beys2019, fujipreprint}. 
{Using the RHS of Eq.~(\ref{eqn:equL}) and 
taking} the shear to be $U/r_{\rm tube}$, 
where $U$ is defined in the penultimate paragraph of
Section \ref{sec:dis}, {we find that} the suppression does not occur when $U$ is 
smaller than approximately $100\ \mu$m/s {for} $r_{\rm tube}=0.1\ \mu$m, $\xi=50\ $nm
and $\eta_{\rm s}=2\ $mPa$\cdot$s.  In the presence of PA, {$\xi$ being much smaller near the tube wall,} 
the Onsager coefficients are little affected by the critical enhancement and {hence} by its suppression,
as far as examined in the present study. 
\end{section}



\begin{thebibliography}{99}
\bibitem{cahn}
J. W. Cahn, "Critical point wetting,"
  J.~Chem.~Phys.~{\bf 66},  3667 (1977).
\bibitem{beysens1985}
D. Beysens and D. Est{\`e}ve, "Adsorption
  phenomena at the surface of silica spheres in a binary liquid mixture,"
  Phys.~Rev.~Lett.~{\bf 54},  2123  (1985).
\bibitem{beysens1982}
D. Beysens and S. Leibler, "Observation of an
  anomalous adsorption in a critical binary mixture," J. Physique Lett.~{\bf 43}, 133--136  (1982).
\bibitem{bonn} D. Bonn and D. Ross, "Wetting transitions," Rep.~Prog.~Phys.~{\bf 64} 1085--1163 (2001).
\bibitem{binder}
M. N. Binder, {\it Phase Transitions and Critical
  Phenomena VIIIV\/}, Critical behavior at surfaces.  (Academic,
  London, 1983).
\bibitem{okafujiko}
R. Okamoto, Y. Fujitani, and S. Komura, 
  "Drag coefficient of a rigid spherical particle in a near-critical binary
  fluid mixture,"  J.~Phys.~Soc.~Jpn  {\bf 82},  084003  (2013).
\bibitem{furu}  A. Furukawa, A. Gambassi, S. Dietrich, and H. Tanaka, "Nonequilibrium critical Casimir effect in binary fluids,"
Phys.~Rev.~Lett.~{\bf 111}, 055701 (2013).
\bibitem{onukibook}  A. Onuki, {\it Phase Transition Dynamics\/}
  (Cambridge University Press, 2002), \textcolor{black}{Chapter} 6.1.2.  
\bibitem{kawasaki}  K. Kawasaki, "Kinetic equations and time correlation functions of critical fluctuations," Ann.~Phys.~(N.Y.) {\bf 61}, 1 (1970)
\bibitem{ohta} T. Ohta, "Selfconsistent calculation of dynamic critical exponents for classical liquid,"
Prog.~Theor.~Phys.~{\bf 54}, 1566 (1975).
\bibitem{fisher-auyang} M. E. Fisher, and H. Au-Yang,  "Critical wall perturbations and a local free energy functional,"
Physica A~{\bf 101}, 255 (1980).

  \bibitem{OkamotoOnuki} R. Okamoto and A. Onuki, 
"Casimir amplitude and capillary condensation of near-critical binary fluids between parallel plates: Renormalized local functional theory,"
  J. Chem.~Phys.~{\bf 136}, 114704 (2012).
\bibitem{undul}  Y. Fujitani, "Undulation amplitude of a fluid membrane
 in a near-critical binary fluid mixture calculated beyond the Gaussian model
 supposing weak preferential attraction," J. Phys.~Soc.~Jpn. {\bf 86} 044602 (2017).
\bibitem{yabuokaon} S. Yabunaka, R. Okamoto, and A. Onuki, "Hydrodynamics in bridging and aggregation of two colloidal particles in a
  near-critical binary mixture," Soft Matter {\bf 11}, 5738 (2015).
\bibitem{yabufuji} S. Yabunaka and Y. Fujitani,
"Drag coefficient of a rigid spherical particle in a near-critical binary fluid mixture, beyond the regime of the Gaussian model,"
J. Fluid Mech.~{\bf 886} A2 (2020).
\bibitem{gromaz} S. R. de Groot and G. Mazur, {\it Non-equilibrium thermodynamics\/}, (Dover, New York, 1984).
\textcolor{black}{Section IV and Chapter XV}.
\bibitem{roij} S. Samin and R. van Roij, "Interplay between adsorption and hydrodynamics in nanomechanics: Towards tunable membranes,"
Phys. Rev. Lett. {\bf 118}, 014502 (2017).
\bibitem{wolynes} P. G. Wolynes, "Osmotic effects near the critical point," J. Phys. Chem. {\bf 80},  1570--1572 (1976).
\bibitem{xu}
X. Xu and T. Qian, "Generalized Lorentz reciprocal theorem in complex fluids and in nonisothermal
systems," J. Phys.: Condens. Matter {\bf 31}, 475101 (2019).
\bibitem{uematsu} Y. Uematsu and T. Araki, "Electro-osmotic flow of semidilute polyelectrolyte solutions," J. Chem.~Phys.~{\bf 139}, 094901 (2013).
\bibitem{osmflow} C. Lee, C. Cottin-Bizonne, A.-L. Biance, P. Joseph, L. Bocquet, and C. Ybert,
"Osmotic flow through fully permeable nanochannels,"  Phys.~Rev.~Lett.~{\bf 112}, 244501 (2014). 
\bibitem{keh} H. J. Keh,  "Diffusiophoresis of charged particles and diffusioosmosis of
electrolyte solutions," Curr.~Opin.~Coll.~Interf.~Sci.~{\bf 24}, 13--22 (2016).
{\bibitem{haq2}M. Atlas, Rizwan Ul Haq, and T. Mekkaoui, "Active and zero flux of nanoparticles between a squeezing channel with thermal radiation effects," J.~Mol.~Liq. {\bf  223}, 289 (2016).}
\bibitem{veleg} \textcolor{black}{D. Velegol, A. Garg, R. Gusha, A. Kar, and M. Kumar, ``Origins of concentration gradients for diffusiophoresis'', 
Soft Matter {\bf 12}, 4686 (2016).}
{\bibitem{haq1}M. Hamid, M. Usman, Z.H. Khan, R. U. Haq, and W. Wang, "Numerical study of unsteady MHD flow of Williamson nanofluid in a permeable channel with heat source/sink and thermal radiation," Eur.~Phys.~J.~Plus~{\bf  133}, 527 (2018).}
\bibitem{marb} S. Marbach and L. Bocquet, "Osmosis, from molecular insights to large-scale
applications," Chem.~Soc.~Rev.~{\bf 48}, 3102--3144 (2019).
\bibitem{shin} S. Shin, "Diffusiophoretic separation of colloids in microfluidic flows," Phys.~Fluids {\bf 32}, 101302 (2020).
\bibitem{werk} B. L. Werkhoven and R. van Roij, "Coupled water, charge and salt transport in
heterogeneous nano-fluidic systems," Soft Matter {\bf 16}, 1527--1537 (2020).
{\bibitem{siva} V. S. Sivasankar, S. A. Etha, H. S. Sachar, and  S. Das,
``Ionic diffusioosmotic transport in nanochannels grafted with pH-responsive polyelectrolyte brushes modeled using augmented strong stretching theory,'' Phys.~Fluids {\bf 32}, 042003 (2020).}
 {\bibitem{frenkel}S. Ram{\' i}rez-Hinestrosa and D. Frenkel, ``Challenges in modelling diffusiophoretic transport,''
Eur.~Phys.~J. B {\bf 94}, 199 (2021).}
\bibitem{sun} C. Sun, R. Zhou,  Z. Zhao, and B. Bai, "Extending the classical continuum theory to describe water flow
through two-dimensional nanopores,"  Langmuir {\bf 37}, 6158--6167 (2021).
\bibitem{derja} B. V. Derjaguin, S. S. Dukhin, and V. V. Koptelova,  "Capillary osmosis through porous partitions and properties of
boundary layers of solutions," J. Coll.~Interf.~Sci.~{\bf 38}, 584--595 (1972).
\bibitem{anders} J. L. Anderson, "Colloid transport by interfacial forces," Ann.~Rev.~Fluid Mech.~{\bf 21}, 61--99 (1989).
\bibitem{diffphore} Y. Fujitani, ``Diffusiophoresis in a near-critical binary fluid mixture,''
{Phys.~Fluids {\bf 34}, 041701 (2022).}
\bibitem{ons1}  L. Onsager, "Reciprocal relations in irreversible processes I," Phys.~Rev.~{\bf 37}, 405 (1931).
\bibitem{ons2}  L. Onsager, "Reciprocal relations in irreversible processes II," Phys.~Rev.~{\bf 37}, 2265 (1931).
\bibitem{bray}
A. J. Bray and M. A. Moore, "Critical behaviour
  of semi-infinite systems,"  J. Phys.~A: Math.~Gen.~{\bf 10},
  1927--1962 (1977).
\bibitem{diehl86}
H. W. Diehl, {\it Phase Transition and Critical Phenomena
  X\/}, Field theoretical approach to critical behavior at surfaces.
  (Academic, London, 1986).
\bibitem{diehl97}
H. W. Diehl, "The theory of boundary critical
  phenomena,"  Int.~J. Mod.~Phys.~B  {\bf 11}, 3503--3523 (1997).
\bibitem{okaonPRE}
R. Okamoto and A. Onuki, "Attractive interaction
  and bridging transition between neutral colloidal particles due to
  preferential adsorption in a near-critical binary mixture,"  Phys.~Rev.~E
  {\bf 88},  022309 (2013).
\bibitem{Yabu-On}
S. Yabunaka and A. Onuki, "Critical adsorption
  profiles around a sphere and a cylinder in a fluid at criticality: Local
  functional theory,"  Phys.~Rev.~E  {\bf 96},  032127  (2017).
\bibitem{peli}
A. Pelisetto and E. Vicari, "Critical phenomena
  and renormalization-group theory,"  Phys.~Rep.~{\bf 368},  549 (2002).
\bibitem{iwan} I. Iwanowski, K. Leluk, M. Rudowski, and U. Kaatze, 
"Critical dynamics of the binary system nitroethane/3-methylpentane: Relaxation rate and
scaling function," J. Phys.~Chem.~A {\bf 110}, 4313 (2006).
\bibitem{bergmold}
R. F. Berg and M. R. Moldover, 
"Critical exponent for the viscosity of four binary liquids,"  J. Chem.~Phys.
{\bf 89}, 3694--3704 (1989). 
\bibitem{bergmold2} R. F. Berg and M. R. Moldover, 
"Critical exponent for  viscosity,"  Phys.~Rev.~A
{\bf 42}, 7183--7187 (1990). 
\bibitem{sengers} J. V. Sengers, "Transport properties near critical points," Int. J. Thermophys. {\bf 6}, 203--232 (1985).
\bibitem{mirz} S. Z. Mirzaev, R. Behrends, T. Heimburg, J. Haller, and U. Kaatze, 
"Critical behavior of 2,6-dimethylpyridine-water: Measurements of specific heat, dynamic light scattering, and shear viscosity," 
J. Chem. Phys. {\bf 124} 144517 (2006).
\bibitem{stein} A. Stein, S. J. Davidson, J. C. Allegra, and G. F. Allen, "Tracer Diffusion and Shear Viscosity for the System 2,6-Lutidine-Water near the
Lower Critical Point," J. Chem. Phys. {\bf 56} 6164 (1972).
\bibitem{gratt} C. A. Grattoni, R. A. Dawe, C. Y. Seah, and J. D. Gray, 
"Lower Critical Solution Coexistence Curve and Physical Properties
(Density, Viscosity, Surface Tension, and Interfacial Tension) of
2,6-Lutidine $+$ Water," Chem. Eng. Data, {\bf 38}, 516--519 (1993).
\bibitem{jaya} Y. Jayalakshmi, J. S. Van Duijneveldt, and D. Beysens,
"Behavior of density and refractive index in mixtures of 2,6-lutidine and water,"  J. Chem.~Phys.~{\bf 100}  604--609 (1994).
\bibitem{leis} H. M. Leister, J. C. Allegra, and G. F. Allen, 
"Tracer diffusion and shear viscosity in the liquid-liquid critical region," J. Chem.~Phys.~{\bf 51} 3701 (1969).
\bibitem{reeder} J. Reeder, T. E. Block, and C. M. Knobler, "Excess volumes of nitroethane +
3-methylpentane," J. Chem.~Thermodyn.~{\bf 8}, 133 (1976).
\bibitem{swinney} H. L. Swinney and D. L. Henry, "Dynamics of Fluids near the Critical Point: Decay Rate of Order-Parameter Fluctuations,"
Phys.~Rev.~A, {\bf 8}, 2586--2617 (1973).
\bibitem{mood} N. Sharifi-Mood, J. Koplik, and C. Maldarelli, "Molecular Dynamics Simulation of the Motion of Colloidal Nanoparticles in a
Solute Concentration Gradient and a Comparison to the Continuum Limit," Phys. Rev.~Lett.~{\bf 111}, 184501 (2013).
\bibitem{miya} K. Miyazaki, D, Bedeaux, and K. Kitahara, "Nonequilibrium thermodynamics
of multicomponent systems,"  Physica A {\bf 230}, 600 (1996).
\bibitem{gard} C. W. Gardiner,  {\it Handbook of Stochastic Methods\/}, (Springer, Berlin, 1985).
\textcolor{black}{Section} 5.3.
\bibitem{capcondens} S. Yabunaka, R. Okamoto, and A. Onuki,
 "Phase separation in a binary mixture confined between symmetric parallel plates: capillary condensation transition near the bulk critical point", 
Phys.~Rev.~E {\bf 87}, 032405 (2013).
\bibitem{tsori2007} Y. Tsori and L. Leibler, "Phase-separation in ion-containing mixtures in electric fields",  Proc.~Nat.~Acad.~Sci.~{\bf 104},
7348--7350 (2007).  
\bibitem{tsori2011}  S. Samin and Y. Tsori, "Attraction between like-charge surfaces in polar mixtures", Europhys.~Lett.~{\bf 95} 36002 (2011). 
\bibitem{tsori2016} S. Samin and Y. Tsori,  "Reversible pore-gating in aqueous mixtures via external potential", Coll.~Interf.~Sci.~Comm,
{\bf 12}, 9 (2016)
\bibitem{precip} R. Okamoto and A. Onuki, "Precipitation in aqueous mixtures with addition of a strongly hydrophilic or hydrophobic solute", 
Phys.~Rev.~E {\bf 82}, 051501 (2010).
\bibitem{precip2} R. Okamoto and A. Onuki, "Charged colloids in an aqueous mixture with a salt", 
Phys.~Rev.~E {\bf 84}, 051401 (2011).
\bibitem{DJ}  H. W. Diehl and H. K. Janssen, "Boundary conditions for field theory of dynamic critical behavior in semi-infinite systems with
conserved order parameter,"  Phys.~Rev.~A {\bf 45}, 7145 (1992).
\bibitem{effvis}
Y. Fujitani, "Effective viscosity of a near-critical
  binary fluid mixture with colloidal particles dispersed dilutely under weak
  shear,"  J.~Phys.~Soc.~Jpn.~{\bf 83},  084401 (2014).
\bibitem{EPJ} Y. Fujitani, ``Relaxation rate of the shape fluctuation of a fluid membrane immersed in a near-critical binary fluid mixture,''
 Eur.~Phys.~J. E {\bf 39} 31 (2016). 
\bibitem{to} K. To, "Coexixtence curve exponent of a binary mixture with a high molecular weight polymer,"
Phys.~Rev.~E. {\bf 63}, 026108 (2001). 
\bibitem{wims} A. M. Wims, D. Mcintyre, and F. Hynne,
 "Coexistence curve for 3-methylpentane-nitroethane near the critical point," J. Chem.~Phys.~{\bf 50} 616 (1969).
  {\bibitem{fisher-degennes} M. E. Fisher and P. G. de Gennes, "Ph\'{e}nom\`{e}nes aux parois dans un m\'{e}lange binaire critique," C. R. Acad. Sci. Paris B~{\bf  287}, 207 (1978).}
\bibitem{sighalhoh}
E. D. Siggia, P. C. Hohenberg, and B. I. Halperin, 
"Renormalization-group treatment of the critical dynamics of the binary-fluid and gas-liquid transitions,"  Phys.~Rev.~B  {\bf 13},
2110--2123 (1976).
\bibitem{folkmos} R. Folk
and G. Moser, "Critical dynamics: a field-theoretical approach," J. Phys.~A: Math.~Gen.~{\bf 39}, R207--R313 (2006).
\bibitem{onukishear} A. Onuki and K. Kawasaki,
"Nonequilibrium steady state of critical fluids under shear flow: a renormalization group approach," 
Ann.~Phys.~(N.Y.) {\bf 121}, 456--528 (1979).
\bibitem{ohta3} T. Ohta, "Multiplicative renormalization of the anomalous shear viscosity
in classical liquids,"  J. Phys.~C: Solid State Phys.~{\bf 10}, 791 (1977).
\bibitem{bhatt} J. K. Bhattacharjee, R. A. Ferrell, R. S. Basu, and J. V. Sengers, 
"Crossover function for the critical viscosity of a classical fluid," Phys. Rev. A, {\bf 24}, 1469 (1981).
\bibitem{tsai}  B. C. Tsai and D. McIntyre, "Shear viscosity of nitroethane-3-methylpentane in the critical region,"
J.~Chem.~Phys.~{\bf 60}, 937 (1974).
\bibitem{onukirev} A. Onuki, "Phase transitions of fluids in shear flow,"
J. Phys, Condensed Matter {\bf 9}, 6119--6157 (1997).
\bibitem{beys2019}
\textcolor{black}{D. Beysens, ``Brownian motion in strongly
  fluctuating liquid,''.  Thermodynamics of Interfaces and Fluid Mechanics
  {\bf 3},  1--8 (2019).}
\bibitem{fujipreprint} Y. Fujitani, "Suppression of viscosity enhancement around a Brownian particle in a near-critical binary fluid
mixture,"  J. Fluid Mech.~{\bf 907}, A21 (2021). 
\end{thebibliography}
\end{document}